\documentclass[aps,prc,reprint,superscriptaddress]{revtex4-2}
\usepackage{ulem}
\usepackage{amsmath}

\usepackage{url}
\usepackage[colorlinks=true,linkcolor=blue,filecolor=blue, urlcolor=blue,citecolor=blue]{hyperref}
\usepackage{float}
\usepackage[T1]{fontenc}
\usepackage{graphicx}
\usepackage{mathrsfs}
\usepackage{dcolumn}
\usepackage{bm}
\usepackage{cases}
\usepackage{multirow,booktabs}
\usepackage{makecell}
\usepackage{amssymb}
\usepackage{eucal}
\usepackage{ragged2e}
\usepackage[justification=RaggedRight,singlelinecheck=false,format=plain]{caption}
\DeclareCaptionJustification{justified}{\justifying}
\usepackage{natbib}
\usepackage{amsthm,amsmath,amssymb}
\usepackage{mathrsfs}
\usepackage{subeqnarray}
\usepackage{color}
\usepackage{ulem}
\usepackage{soulutf8}
\usepackage{appendix}
\usepackage{subcaption}

\begin{document}
	
	
	\title{Nucleon momentum distributions of complex nuclei from inclusive electron scattering}
	
	
	\author{Tongqi Liang}
	\email{liangtq@ouc.edu.cn}
	\affiliation{College of Physics and Optoelectric Engineering, Ocean University of China, Qingdao 266100, China}

	 \author{Dong Bai}
	 \affiliation{College of Mechanics and Engineering Science, Hohai University, Nanjing 211100, Jiangsu, China}
	
	\author{Zhongzhou Ren}
	\affiliation{School of Physics Science and Engineering, Tongji University, Shanghai 200092, China}
	\affiliation{Key Laboratory of Advanced Micro-Structure Materials, Ministry of Education, Shanghai 200092, China}

	
	
	\begin{abstract}
Nucleon momentum distributions (NMDs) reveal essential information about Fermi motion and short-range correlations (SRCs).
In extracting NMDs from inclusive electron scattering data, theoretical analyses, such as the scaling analysis, are typically employed.
For complex nuclei, consistently treating the excitation energy of the residual system is a complicated task, leading to discrepancies between existing extracted NMDs and ab initio calculations, particularly around the Fermi momentum $k_F$. 
To address this issue, we introduce an improved description of the excitation energy in the framework of the relativistic Fermi gas (RFG) model.
With this treatment, the extracted NMDs of complex nuclei show better agreement with ab initio calculations across the low- and high-momentum range, especially around $k_F$, successfully reproducing both the behaviors of Fermi motion and SRCs.
These results provide a new experimental perspective on the interplay between Fermi motion and SRCs in complex nuclei.



	\end{abstract}

	\maketitle
	
	\section{ introduction}

The independent particle model (IPM) has long served as a fundamental framework for describing nuclear structure and reactions~\cite{Sarriguren2019,Sarriguren2019tqu,Reinhard2021,Stoitsov2004,SCHUNCK2017}.
Within this picture, nucleons move almost independently in a mean-field potential, and the model successfully reproduces the momentum distribution of nucleons with momenta below the Fermi momentum $k_F$~\cite{Wang2021,MoyadeGuerra1991}.
However, the IPM fails to account for the high-momentum components of nucleon momentum distributions (NMDs), which arise from nucleon–nucleon ($NN$) short-range correlations (SRCs).
Extensive experimental and theoretical studies have firmly established the existence of SRCs, revealing that about 20–25\% of nucleons in medium and heavy nuclei participate in correlated pairs~\cite{Li2025,Qin2023,Niu2022,Duer2019,Hen2014,Cruz-Torres2019,Weiss2016}.
A comprehensive and unified description of both low- and high-momentum nucleons is essential for a consistent understanding of nuclear dynamics and the underlying $NN$ interactions~\cite{Hen2012,Hen2016,Schmookler2019,Hong2024,Liang20242,Liang2025,Chen2017,Jokiniemi2021,Weiss2022,Li20252,Lu2022,Cai2025}.

Over the past decades, many theoretical approaches have been developed to calculate NMDs, including quantum Monte Carlo (QMC), correlated basis function (CBF), and no core shell model (NCSM), which have provided valuable microscopic insight into NMDs~\cite{Wiringa2014,Piarulli2022,Marcucci2019,Tropiano2021,Neff2015,Tropiano2024,Meng2023,Li2024}. 
Experimentally, inclusive ($e,e'$) electron scattering has proven to be a clean and powerful probe of NMDs and SRCs~\cite{Li2022,Ivanov2020,Day1990,Caballero2010,Wang2023,Liang2022,Fomin2012,Arrington2022,Schmidt2024,Zhang2025}.
At sufficiently high momentum transfers, the plane wave impulse approximation (PWIA) offers a reliable framework for interpreting the inclusive cross sections~\cite{Ciofi2015}.
In such an approximation, the electron interacts with a bound, off-shell nucleon, and the nucleon is knocked out from the nucleus, leaving the residual nuclear system in an excited state characterized by its excitation energy.
The excitation energy thus plays a key role in connecting the inclusive cross section with the underlying momentum distribution.


Under the PWIA framework, the inclusive electron cross section can be expressed as a product of the single-nucleon cross section and a scaling function, which is closely related to the longitudinal momentum distribution.
The $y$-scaling analysis has been widely employed to analyze inclusive cross sections and extract information about NMDs~\cite{Ciofi1991,Ciofi1999,Ciofi2009}.
Despite its success, noticeable discrepancies remain between theoretical predictions and experimentally extracted NMDs, especially around the Fermi momentum, the transition region between Fermi motion and SRCs. 
These discrepancies may originate from limitations of the theoretical framework or from uncertainties in the extraction procedure.
In particular, the treatment of the excitation energy of the residual nuclear system can significantly affect the reliability of NMDs extracted from inclusive scattering data~\cite{Benhar2013,Ciofi1994}.


In previous $y$-scaling studies, different approximations were adopted for the excitation energy of the residual system. For instance, $y_{\text{CW}}$ treats it as the kinetic energy of the correlated nucleon~\cite{Ciofi2009}. 
Over the past decades, extensive theoretical and experimental efforts have been devoted to understanding the excitation energy~\cite{Rocco2015,Weiss2018,Ankowski2024,Rohe2004,Benmokhtar2004,Schmidt2020,Gu2020,Jiang2023}. 
However, these results are difficult to apply to the extraction of NMDs due to their complexity and non-analytical form.
In this paper, we propose an improved description of the excitation energy within the relativistic Fermi gas (RFG) model, providing a consistent framework that involves the Fermi motion and SRCs.
We analyze this excitation-energy treatment with the $y$-scaling function and extract NMDs from inclusive electron scattering for a broad range of complex nuclei.
The resulting NMDs exhibit expected physical features, simultaneously reproducing the low-momentum plateau associated with Fermi motion and the universal high-momentum tail driven by SRCs.
In our recent studies, the NMDs of the deuteron have been extracted using modern numerical differentiation techniques, with the nucleon momentum $k$ extended to 1.2 GeV/c~\cite{Liang2024}.
However, similar analyses for complex nuclei are hindered by the treatment of excitation energy.
The present work extends such analyses to complex nuclei, enabling a systematic extraction of NMDs over a broad range of nuclei and momenta.
A consistent treatment of the excitation energy of bound nucleons is also important for improving our understanding of both the studies of electron- and neutrino-nuclei scattering processes.

The rest of the paper is organized as follows: In Sec.~\ref{secII}, we present the theoretical framework for extracting NMDs from inclusive electron scattering data and the expression of RFG excitation energy.
In Sec.~\ref{NR}, we present the extractions of NMDs for complex nuclei and address the impact of RFG excitation energy.
Finally, conclusions are drawn in Sec.~\ref{Concl}.

	





	\section{theoretical framework}\label{secII}

 
Under the assumption of PWIA, the inclusive electron–nucleus scattering can be treated as the one–nucleon knockout process, in which the ejected nucleon does not interact with the recoil $A-1$ system \cite{Day1990}. 
In this framework, the inclusive cross section $\frac{\text{d}^2\sigma}{\text{d}\omega\text{d}\Omega}$ can be written as the product of the single-nucleon cross section $\sigma_{ep(en)}$ and the nuclear structure function $F(q, \omega)$,
		\begin{equation}\label{PWIA}
			\frac{\text{d}^2\sigma}{\text{d}\omega\text{d}\Omega}=\left(Z \sigma_{e p}+N \sigma_{e n}\right)\Bigg|\overline{\frac{\partial \omega}{k \,\partial\!\cos \alpha}}\Bigg|^{-1} F(q, \omega).
		\end{equation}
Here, $Q^2=q^2-\omega^2$, with $\omega$ and $q=|\textbf{q}|$ the energy transfer and the momentum transfer of the electron.
$Z$ and $N$ are the numbers of protons and neutrons of the target nucleus, and $\left|\overline{\frac{\partial \omega}{k \,\partial\!\cos \alpha}}\right|^{-1}$ (with $\cos \alpha=\frac{\textbf{q}\cdot \textbf{k}}{qk}$) is the kinematic factor arising from the energy conservation in the scattering process. 
The single-nucleon cross sections $\sigma_{ep}$ and $\sigma_{en}$ are taken to be $\sigma_{CC1}$~\cite{Forest1983} with the parameterizations of the elastic form factors as in Refs.~\cite{Arrington2007,Kelly2004}.

The nuclear structure function $F(q,\omega)$ is the integral of the nucleon spectral function ${P_N}(k,E)$
		\begin{equation}\label{Fqw}
			F(q,\omega)=2\pi{\int_{{E_{\min }}}^{{E_{\max }}(q,\omega )} \mathrm{d} } E\int_{{k_{\min }}(q,\omega ,E)}^{{k_{\max }}(q,\omega ,E)} k{\kern 1pt} {\kern 1pt} {P_N}(k,E) \,\mathrm{d} k{\kern 1pt} {\kern 1pt} {\kern 1pt} ,
		\end{equation}
Here, $E_{\text{min(max)}}$ and $k_{\text{min(max)}}$ denote the minimum (maximum) removal energy and momentum accessible to a nucleon when it is knocked out of the nucleus by the electron. 
The nucleon spectral function $P_N(k,E)$ characterizes the joint probability of finding a nucleon in the nucleus with momentum $k$ and removal energy $E$.

\subsection{Scaling analysis}
		
Assuming that the recoil $A-1$ system is not excited, i.e. $E=E_{\text{min}}$, $k_{\min }$ is renamed as the scaling variable $y$, which only depends on $q$ and $\omega$
		\begin{equation}
			|y|=k_{\text{min}}(q,\omega,E_{\text{min}}).
		\end{equation}
The scaling variable $y$ then can be determined
		\begin{eqnarray}\label{energy_2}
			\omega  \!+\! {M_A} \!=\! {\left[ {{m_N^2} \!+\! {{(q+y)}^2}} \right]^{1/2}} \!+\! {\left( {M_{A - 1}^{2} \!+\! y^2} \right)^{1/2}},
		\end{eqnarray}
where $m_N$, ${M_A}$, and $M_{A - 1}$ denote the masses of the knocked-out nucleon, the target nucleus and the recoil system.

Since the spectral function $P_N(k,E)$ falls off rapidly with increasing $k$ and $E$, the upper limits $E_{\text{max}}$ and $k_{\text{max}}$ in Eq.~(\ref{Fqw}) can be safely extended to infinity. In the limit of large $Q^{2}$, the nuclear structure function $F(q,\omega)$ reduces to the scaling function $F(y)$, which only depends on the scaling variable $y$
		\begin{equation}\label{strucFy2}
			F(y)=2\pi{\int_{{E_{\min }}}^{\infty} \mathrm{d} } E\int_{|y|}^{\infty} {\kern 1pt} {\kern 1pt} {\kern 1pt} k{\kern 1pt} {\kern 1pt} {P_N}(k,E)\,\mathrm{d} k.
		\end{equation}

For the deuteron, the recoil system has no excitation energy, and the removal energy $E$ reduces to the minimal separation energy $E_{\min}=2.225$~MeV. In this case, the effects of excitation energy can be safely neglected.
The spectral function takes the approximate form
$P_N(k,E) = n(k)\,\delta(E - E_{\min}),$
and is therefore entirely determined by the nucleon momentum distribution $n(k)$.
The scaling function $F(y)$ becomes
		\begin{equation}\label{Fy}
			F(y)=2\pi\int_{|y|}^{\infty}n(k)\,\,k\,\,\mathrm{d} k,
		\end{equation}
where $n(k)=\int_{E_{\text{min}}}^{\infty}P_N(k,E)\,\mathrm{d}E$. 
Then one can extract the nucleon momentum distribution $n(k)$ from the $y$-scaling function,
		\begin{align}
			n(k)=-\frac{1}{2\pi y}\frac{\mathrm{d}F(y)}{\mathrm{d}y}\bigg|_{\mid y\mid=k},
		\end{align}
By utilizing Eq.~\eqref{PWIA}, $F(y)$ is extracted from the inclusive cross sections.

The normalization of the $y$-scaling function is given by~\cite{Ciofi1999}
\begin{equation}
    \int n(k)\, \text{d}^3k=\int_{-\infty}^{\infty} f(y) \,\text{d}y =1.
\end{equation}\label{normalization}

\subsection{RFG excitation energy}


\begin{figure*}[t] \centering
\includegraphics[width=1.7\columnwidth]{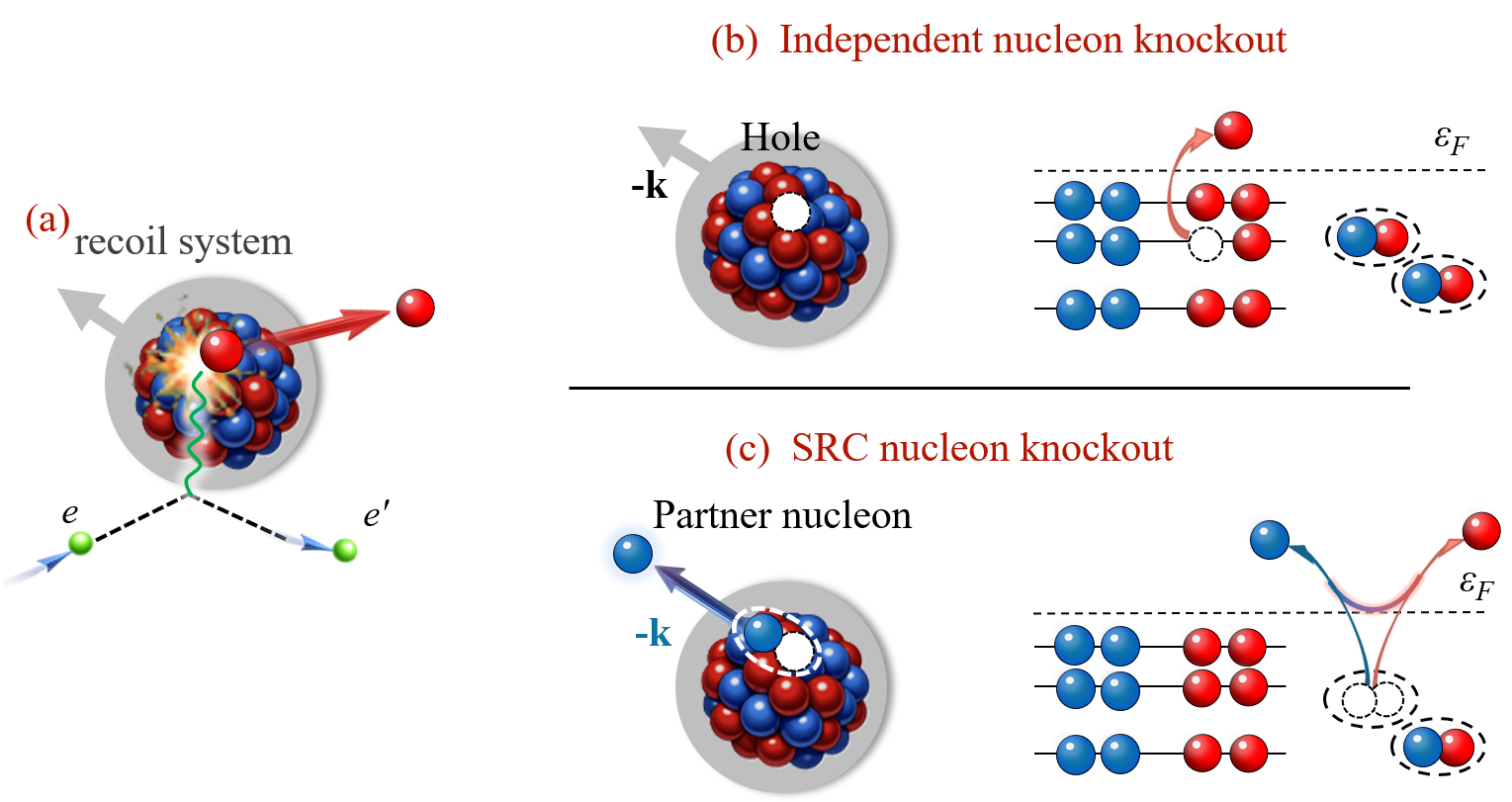}
		\caption{(a) Illustration of the one-nucleon knockout process in electron-nucleus scattering under the PWIA assumption. 
		The residual $A-1$ system is left in an excited state after a nucleon with momentum $\bf{k}$ is knocked out.
		For (b) independent nucleon knockout, removing a nucleon creates a single hole below the Fermi energy $\varepsilon_F$, and the $A-1$ nucleus recoils with corresponding $-\bf{k}$.
		For (c) SRC nucleon knockout process, its partner nucleon carries an approximately back-to-back momentum $-\bf{k}$, and the remaining $A-2$ system stays nearly at rest.}
		\label{kinematic}
\end{figure*}

As illustrated in Fig.~\ref{kinematic}, removing a nucleon from a complex nucleus leaves the recoil $A-1$ system in an excited state.
Hence, the removal energy differs from, and is typically larger than, the minimal separation energy $E_{\text{min}}$.
For complex nuclei, the expression of the energy conservation is rewritten as 
		\begin{eqnarray}\label{energy_2_complex}
			\omega  \!+\! {M_A} \!=\! {\left[ {{m_N^2} \!+\! {{(q+y)}^2}} \right]^{1/2}} \!+\! {\left( {M^{*2}_{A - 1} \!+\! y^2} \right)^{1/2}},
		\end{eqnarray}
where $M^*_{A - 1}=M_{A - 1}+E^*_{A-1}$ is the mass of the excited residual $A-1$ system, with $M_{A - 1}$ the rest mass and $E^*_{A-1}$ the intrinsic excitation energy.
In this work, we propose a unified, momentum-dependent description for the average excitation energy $E^*_{A-1}$ within the framework of the relativistic Fermi gas (RFG) model.
For independent nucleons with $k<k_F$, the knocked-out nucleon is assumed to occupy a state below the Fermi energy $\varepsilon_F$.
Removing it creates a hole of energy $\sqrt{k^2+m^2_N}$.
Thus the intrinsic excitation energy of the residual $A-1$ nucleus is given by
\begin{equation}\label{lekf}
	E^{*\text{RFG}}_{A-1}=\sqrt{k^2_F+m^2_N}-\sqrt{{k}^2+m^2_N}, \,\,\,k<k_F.
\end{equation}

For $k>k_F$, the struck nucleon predominantly originates from SRC pairs.
In previous $y_{\text{CW}}$-scaling analyses, the average excitation energy of the residual system associated with SRCs was effectively represented by the kinetic energy of the correlated partner with momentum $-\bf{k}$, taken as
$\sqrt{k^2+m^2_N}-m_N$~\cite{Ciofi1999,Ciofi2009}.  
This assumption was further supported by the missing-momentum and energy dependence of one- and two-proton knockout reaction~\cite{Shneor2007,Schmidt2020}.

In the present work, we refine this treatment so that the excitation energy is defined consistently within the RFG framework introduced in Eq.~(\ref{lekf}).
The treatment used in $y_{\text{CW}}$-scaling analyses, when applied directly to the RFG model, results in a discontinuity between the excitation energies on the two sides of $k_F$.
For instance, with $k_F=285$ MeV/c, the excitation energy at $k=k_F$ is $\sqrt{k^{2}+m_{N}^{2}}-m_{N}\approx$ 42 MeV above $k_F$, while it vanishes below $k_F$ according to Eq.~(\ref{lekf}).
It is worth noting that the correlated partner in Eq.~(\ref{energy_2_complex}) is still treated as a constituent inside the $A-1$ system rather than as a free nucleon.
The correlated partner thus should be referenced to the Fermi level rather than to the free-nucleon rest energy.
With these considerations, removing a SRC nucleon with momentum $k>k_F$ leaves the excitation energy
\begin{equation}\label{gekf}
	\sqrt{k^{2}+m_{N}^{2}}-\sqrt{k_{F}^{2}+m_{N}^{2}} .
\end{equation}
With this treatment, the excitation energy vanishes at $k=k_F$ like Eq.~(\ref{lekf}) for $k<k_F$, ensuring continuity across the Fermi momentum.

For $k>k_F$, the expression in Eq.~(\ref{gekf}) contains both the intrinsic excitation energy of the residual system and the recoil kinetic energy of the $A-1$ system $E^{R}_{A-1}=\sqrt{\mathbf{k}^2+M^2_{A-1}}-M_{A-1}$.
Subtracting the recoil part and combining the low- and high-momentum regions yields a unified $E^{*\text{RFG}}_{A-1}$
\begin{equation}\label{totalexcitation}
    E^{*\text{RFG}}_{A-1}=\begin{cases}\sqrt{k^2_F+m^2_N}-\sqrt{\mathbf{k}^2+m^2_N}, & k<k_F, \\ \sqrt{\mathbf{k}^2+m^2_N}-\sqrt{k^2_F+m^2_N}-E^{\text{R}}_{A-1}, & k>k_F,\end{cases}
\end{equation}
where the upper branch smoothly reproduces the mean-field Fermi-motion excitation, and the lower branch incorporates SRC physics together with a consistent relativistic treatment of recoil effects.

	\section{\label{NR}numerical results}

As discussed in the above sections, the scaling analysis is reasonable for the deuteron, and the $n(k)$ extractions have been presented in many works \cite{Liang2024,Fomin2012}.
For complex nuclei, however, there are discrepancies between the extracted $n(k)$ results and ab initio calculations, due to the excitation energy of the residual system.
In the following, we present the $n(k)$ extractions for complex nuclei with the RFG excitation energy in Eq.~(\ref{totalexcitation}), and discuss the validity of this treatment.
In our model, $k_F$ represents the position of zero excitation energy and also the separation between the Fermi motion and SRC regimes.
The values of $k_F$ go from 240 to 300 MeV/c for nuclei from $^3$He to nuclear matter.


\begin{figure}[h]
\centering

\begin{subfigure}{0.9\linewidth}
  \includegraphics[width=\linewidth]{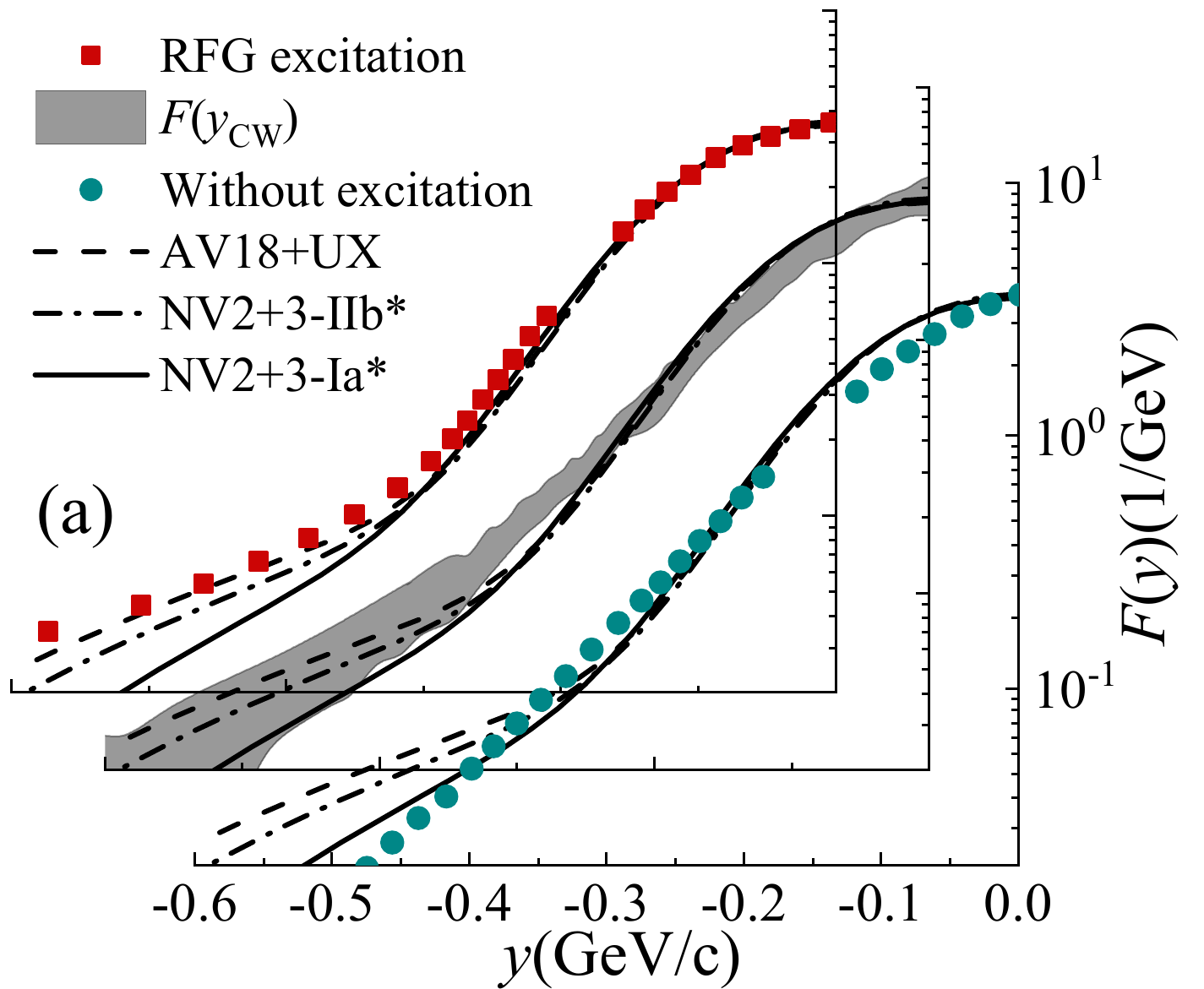}
\end{subfigure}
\hfill
\begin{subfigure}{0.9\linewidth}
  \includegraphics[width=\linewidth]{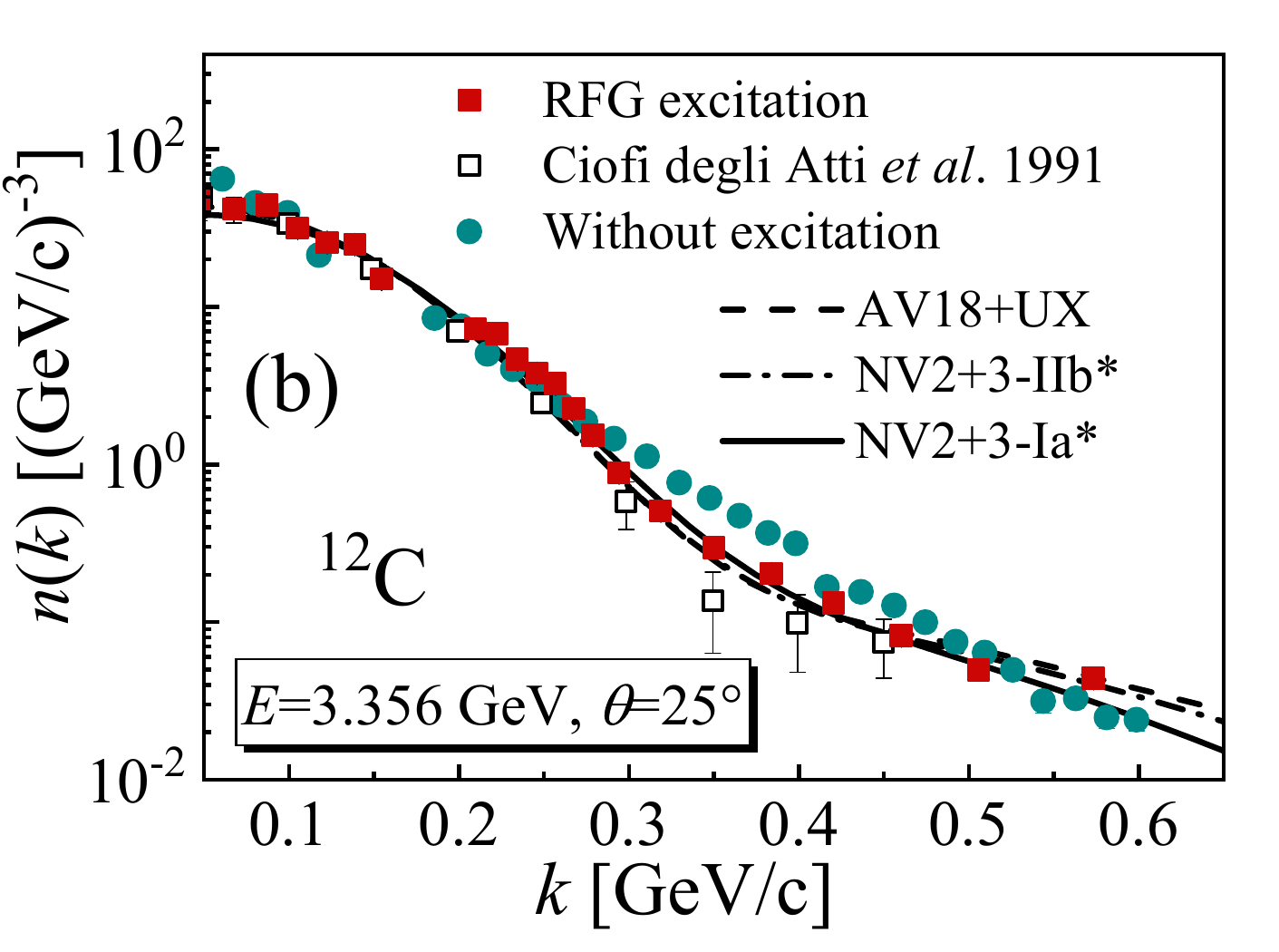}
\end{subfigure}
\caption{(a) $y$-scaling functions $F(y)$ and (b) nucleon momentum distributions $n(k)$ of $^{12}$C extracted from the inclusive cross section with the incident electron energy $E_e=3.356$ GeV and the scattering angle $\theta=25^\circ$~\cite{Zhang2025}. Red squares indicate the extractions with RFG excitation energy, and green circles indicate the extractions without excitation energy. The $F(y_{\text{CW}})$ results (shaded area)~\cite{Ciofi2009}, extractions by Ciofi degli Atti \textit{et al}. (black squares)~\cite{Ciofi1991}, and QMC calculations (lines) with different $NN+3N$ interactions are presented for comparison~\cite{Wiringa2014}.
}
\label{Fy_C12_3356_25}
\end{figure}

Acting as the connection between inclusive cross sections and $n(k)$, the $y$-scaling function $F(y)$ is expressed as the integration of $n(k)$ (as shown in Eq.~(\ref{Fy})).
Therefore, the $F(y)$ results in the low- and high-$|y|$ regions can also reflect the Fermi motion and SRCs, respectively.
Figure~\ref{Fy_C12_3356_25}(a) shows the extracted $y$-scaling function $F(y)$ for $^{12}$C. 
The latest E08-14 experimental data are used with the incident electron energy $E_{e}=3.356\text{ GeV}$ and the scattering angle $\theta=25^{\circ}$~\cite{Zhang2025}.
We present the $F(y)$ extractions with RFG excitation energy, and for comparison, the $F(y)$ extractions without excitation energy, the $F(y_{\text{CW}})$ results, as well as the QMC results with different $NN+3N$ potentials (AV18+UX, NV2+3-Ia*, and NV2+3-IIb).
Notice that the $F(y_{\text{CW}})$ results are taken from Ref.~\cite{Ciofi2009}, which contain several experimental data sets.
Therefore, the $F(y_{\text{CW}})$ results in Fig.~\ref{Fy_C12_3356_25}(a) are presented using the shaded area.
As illustrated in Fig.~\ref{Fy_C12_3356_25}(a), the $F(y)$ results without excitation energy decrease exponentially as $|y|$ increases, which underestimate the QMC results at low-$|y|$ region, and overestimate the QMC results around $|y|\sim$ 0.3 GeV/c. 
Considering the excitation energy as the kinetic energy of the correlated nucleon, the $F(y_{\text{CW}})$ results show a good behavior at high-$|y|$ region, where SRCs dominate.
However, lacking a proper description of the Fermi motion, the $F(y_{\text{CW}})$ results deviate from the QMC results in the region $|y|<0.4$ GeV/c, showing a similar behavior to the one without excitation energy.
The $F(y)$ results with RFG excitation energy show a better agreement with the QMC results in both the low- and high-$|y|$ regions.

To intuitively understand the improvement of the RFG excitation energy, we plot the nucleon momentum distribution $n(k)$ for $^{12}$C in Fig.~\ref{Fy_C12_3356_25}(b), which is extracted from the corresponding $F(y)$ results in Fig.~\ref{Fy_C12_3356_25}(a).
The Gaussian profile, centered around the measured values with a width given by the uncertainty, is used to propagate the uncertainties from $F(y)$ to $n(k)$.
Similar to the $F(y)$ results, the $n(k)$ results without excitation energy fail to describe both the Fermi-motion and SRC behaviors.
The widely used $n(k)$ results in Ref.~\cite{Ciofi1991} are presented for comparison.
With the RFG excitation energy, the extracted $n(k)$ results show a good agreement with the QMC results, successfully reproducing both the Fermi-motion and high-momentum-tail behaviors, especially showing a significant improvement around the transition region.
The improvement reveals the reasonable description of the RFG excitation energy for $^{12}$C.

\begin{figure}[h] \centering
\includegraphics[width=0.8\columnwidth]{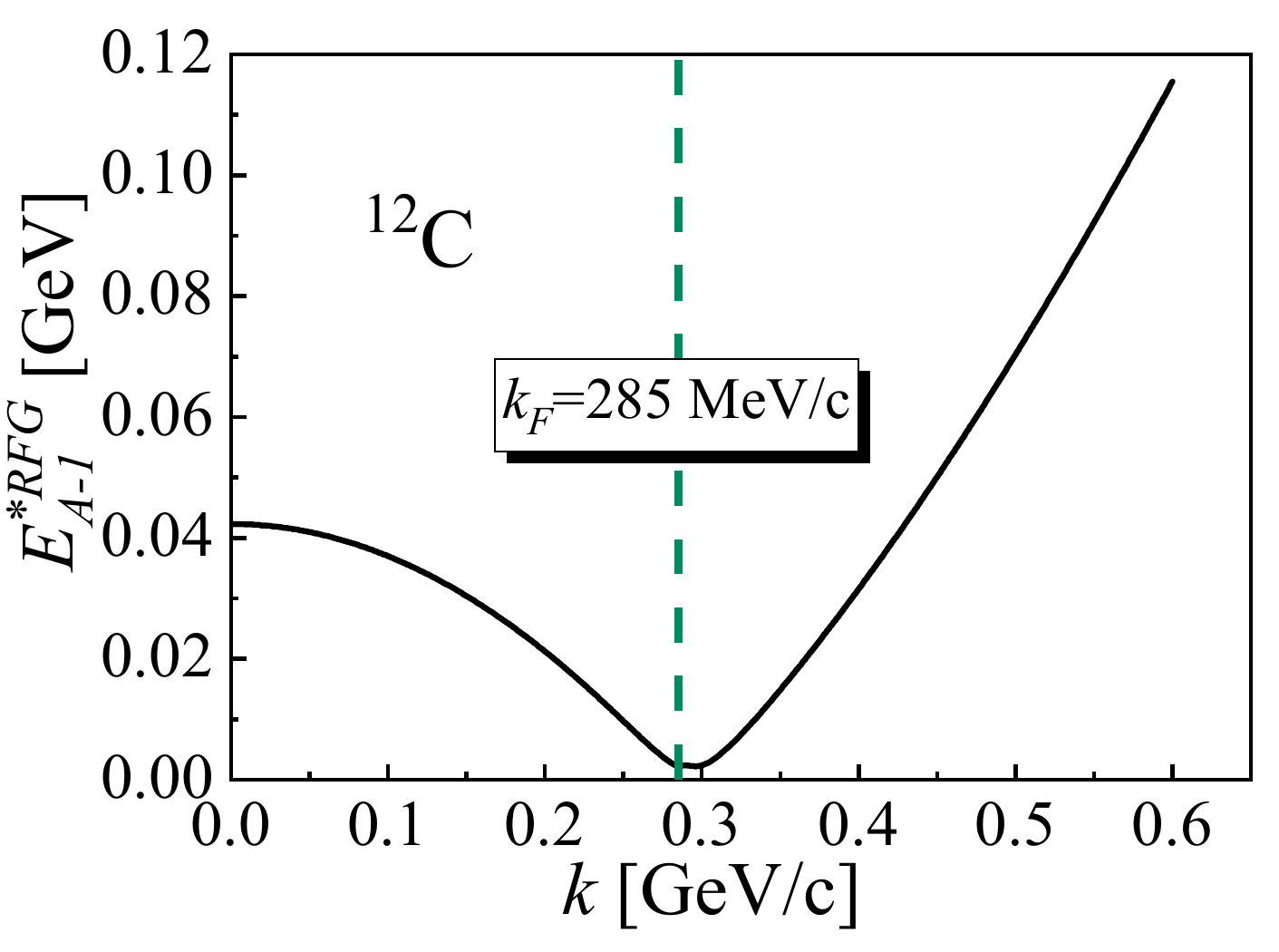}
		\caption{The RFG excitation energy $E^{*\text{RFG}}_{A-1}$ as a function of $k$ for $^{12}$C with the Fermi momentum setting as $k_F=285$ MeV/c.}
		\label{excitation_energy}
\end{figure}

To illustrate how the RFG excitation energy works in the scaling analysis, in Fig.~\ref{excitation_energy}, we present the RFG excitation energy $E^{*\text{RFG}}_{A-1}$ in Eq.~(\ref{totalexcitation}) for $^{12}$C.
Noticing that in Eq.~(\ref{totalexcitation}) the subtraction of $E^{\text{R}}_{A-1}$ induces a small mismatch in $E^{*\text{RFG}}_{A-1}$ across the Fermi momentum.
This difference, however, is numerically modest compared with the overall excitation energy.
For instance, in $^{12}$C with $k_F=285$ MeV/c, the excitation energy above $k_F$ differs from that below $k_F$ by only $\sim4$ MeV after subtracting the recoil energy.
To ensure the continuity at $k=k_F$, we smoothly moderate the $E^{R}_{A-1}$ correction in the $k>k_F$ region by multiplying it with a transition function that varies from 0 to 1 across a narrow momentum window.
A softplus function is then used to blend the two branches, ensuring a smooth first-derivative connection.
As shown in Fig.~\ref{excitation_energy}, the RFG excitation energy decreases with $k$ in the Fermi-motion region ($k<k_F$), while it increases with $k$ in the SRC region ($k>k_F$).
Compared to the traditional $y$ scaling variable without $E^{*\text{RFG}}_{A-1}$, the introduction of $E^{*\text{RFG}}_{A-1}$ shifts the position of $y$ to $k_F$ in the region $k<k_F$.
In the region $k>k_F$, the increasing behavior of $E^{*\text{RFG}}_{A-1}$ with $k$ strengthens the growth rate of $|y|$.
These two features of $E^{*\text{RFG}}_{A-1}$ make the scaling variable $y$ more consistent with the nucleon momentum $k$ in both low- and high-momentum regions, effectively improving the extraction of $F(y)$ and $n(k)$ from inclusive cross sections.

\begin{figure}[h] \centering
\includegraphics[width=0.9\columnwidth]{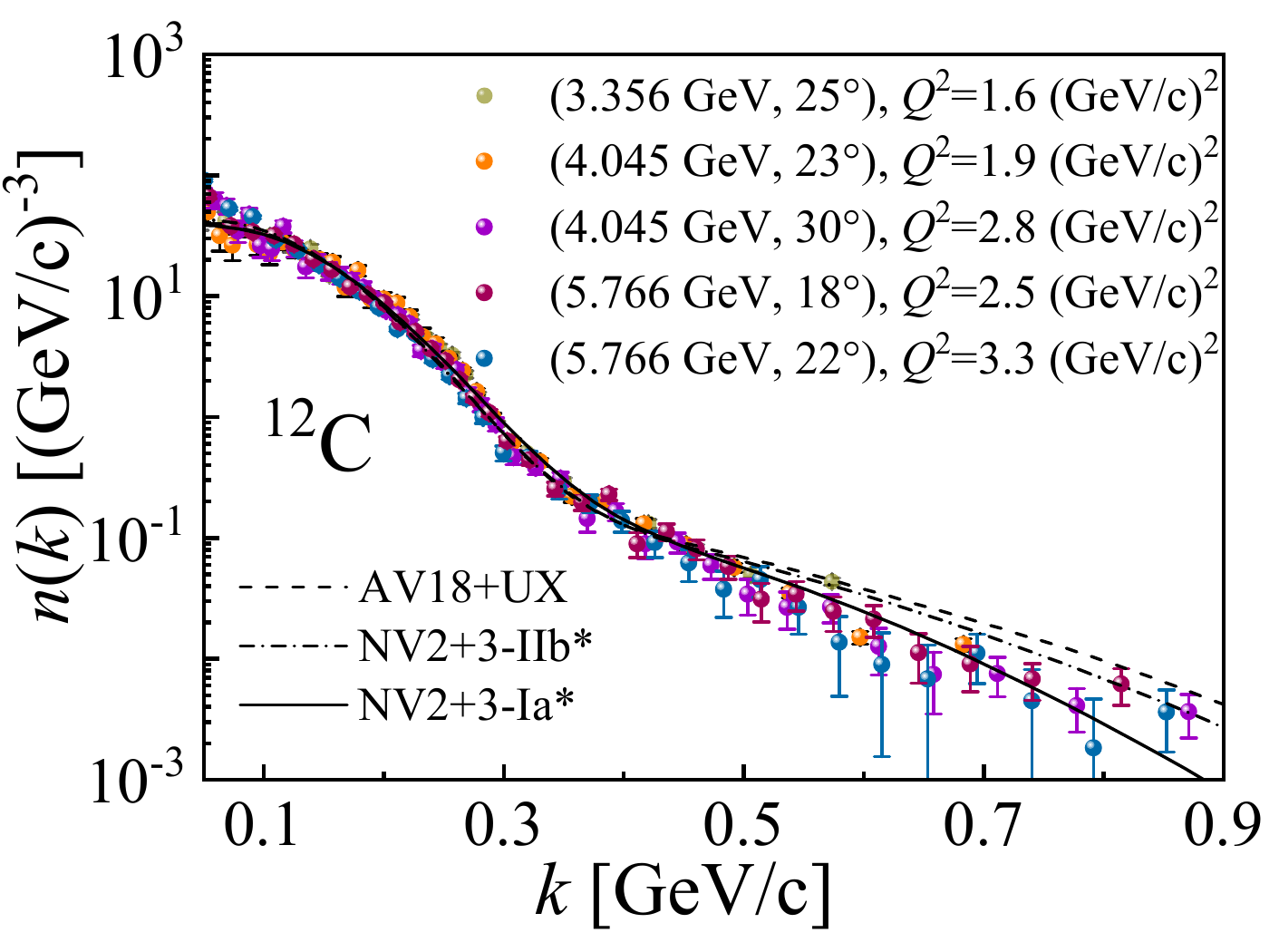}
		\caption{Nucleon momentum distributions $n(k)$ of $^{12}$C extracted from the experimental cross sections measured at ($E_e,\theta$)=(3.356 GeV, 25$^{\circ}$), (4.045 GeV, 23$^{\circ}$), (4.045 GeV, 30$^{\circ}$), (5.766 GeV, 18$^{\circ}$), and (5.766 GeV, 22$^{\circ}$). The $Q^2$ values at the quasielastic peak are also given.}
		\label{nk_C12}
\end{figure}

The dependence on experimental kinematics is examined in Fig.~\ref{nk_C12}, where we present the extracted $n(k)$ results for $^{12}$C with different kinematics.
We present five sets of experimental data from JLab E89-008~\cite{Arrington1999}, E02-019~\cite{Fomin2012} and E08-014~\cite{Ye2013} experiments, with large $Q^2$ values ranging from 1.6 to 3.3 $\text{(GeV/c)}^2$.
Despite spanning a broad range of energy and momentum transfers, these data sets yield a consistent $n(k)$ extraction.
These results exhibit good agreement with the QMC calculations in both the low- and high-momentum regions.
Around the Fermi momentum $k_F$, the extracted $n(k)$ results from different kinematics also show a good consistency, indicating the reliability of the RFG excitation energy treatment in this transition region.
The overall consistency across different kinematics provides strong evidence for the reliability of $E^{*\text{RFG}}_{A-1}$.

\begin{figure}[h] \centering
\includegraphics[width=0.9\columnwidth]{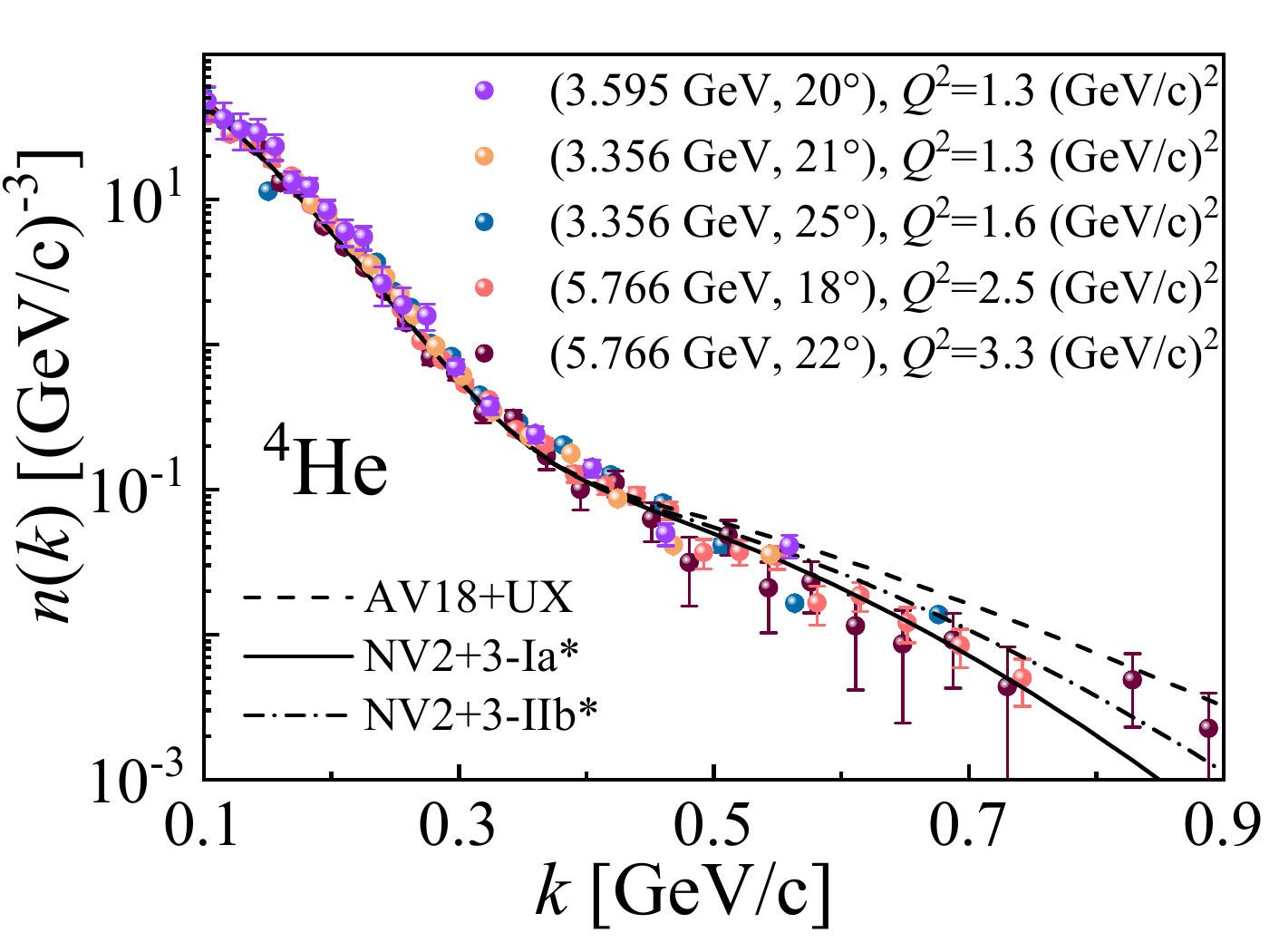}
		\caption{Nucleon momentum distributions $n(k)$ of $^{4}$He extracted from the experimental cross sections measured at ($E_e,\theta$)=(3.595 GeV, 20$^{\circ}$), (3.356 GeV, 21$^{\circ}$), (3.356 GeV, 25$^{\circ}$), (5.766 GeV, 18$^{\circ}$), and (5.766 GeV, 22$^{\circ}$). The $Q^2$ values at the quasielastic peak are also given.}
		\label{nk_He4}
\end{figure}

To further validate the approach, we present the extracted $n(k)$ results for $^{4}$He in Fig.~\ref{nk_He4} under various kinematic conditions.
It is shown in the figure that, with the RFG excitation energy, the extracted nucleon momentum distributions are consistent with the QMC calculations, accurately capturing both the Fermi-motion region and the high-momentum tail.
Moreover, the consistency of $n(k)$ extracted from different experimental data demonstrates a weak dependence on the kinematic settings.
These results support the applicability and reliability of the RFG excitation-energy description.

\begin{figure*}[t]
\centering
\begin{subfigure}{0.333\textwidth}
  \caption{($5.766~\text{GeV},22^{\circ}$), $Q^2=3.3~(\text{GeV/c})^2$}
  \label{nk_H2_5766_18}
  \includegraphics[width=\linewidth]{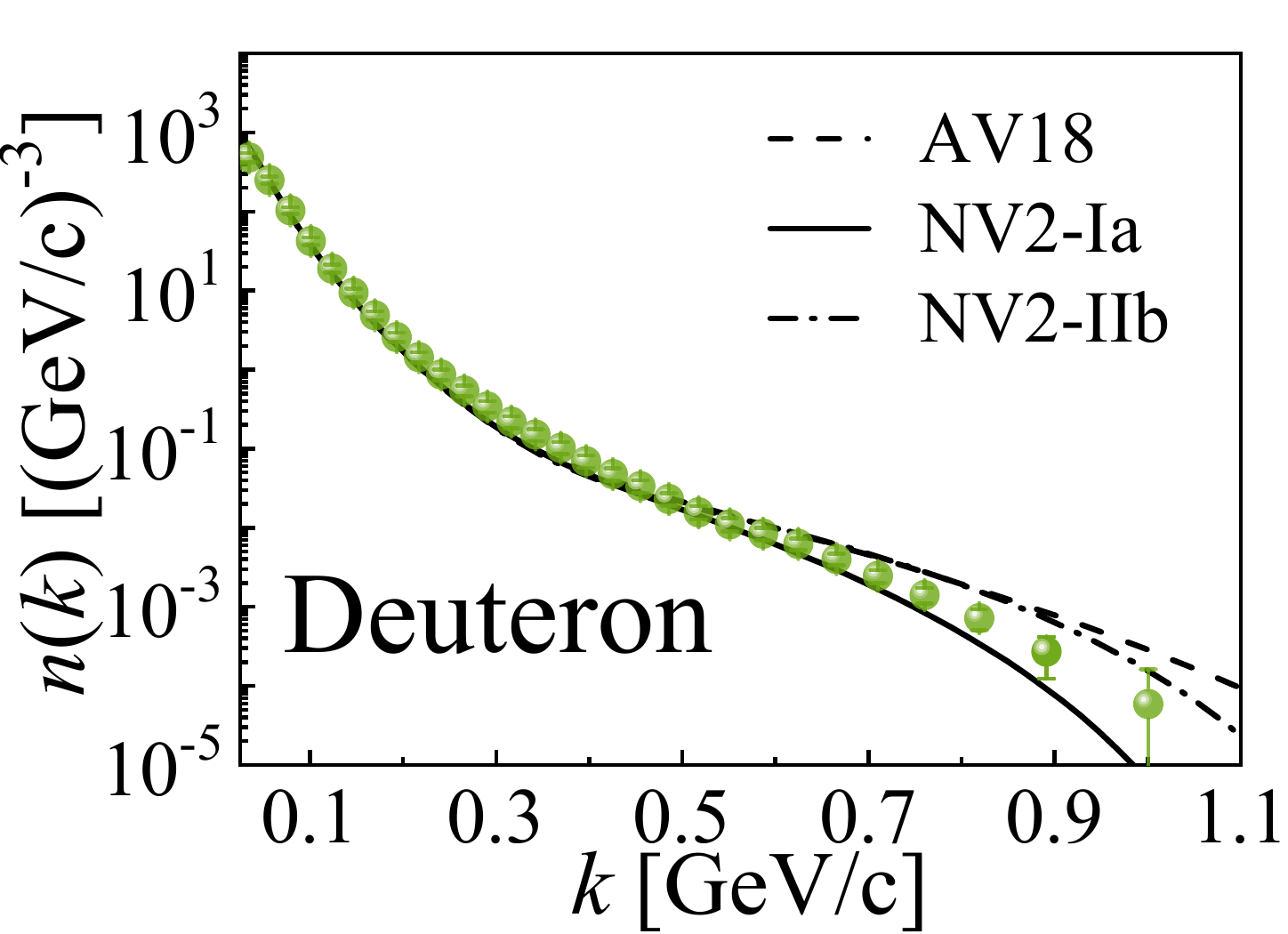}
\end{subfigure}
\hfill
\begin{subfigure}{0.313\textwidth}
  \caption{($5.766~\text{GeV},18^{\circ}$), $Q^2=2.5~(\text{GeV/c})^2$}
  \label{nk_He3_5766_18}
  \includegraphics[width=\linewidth]{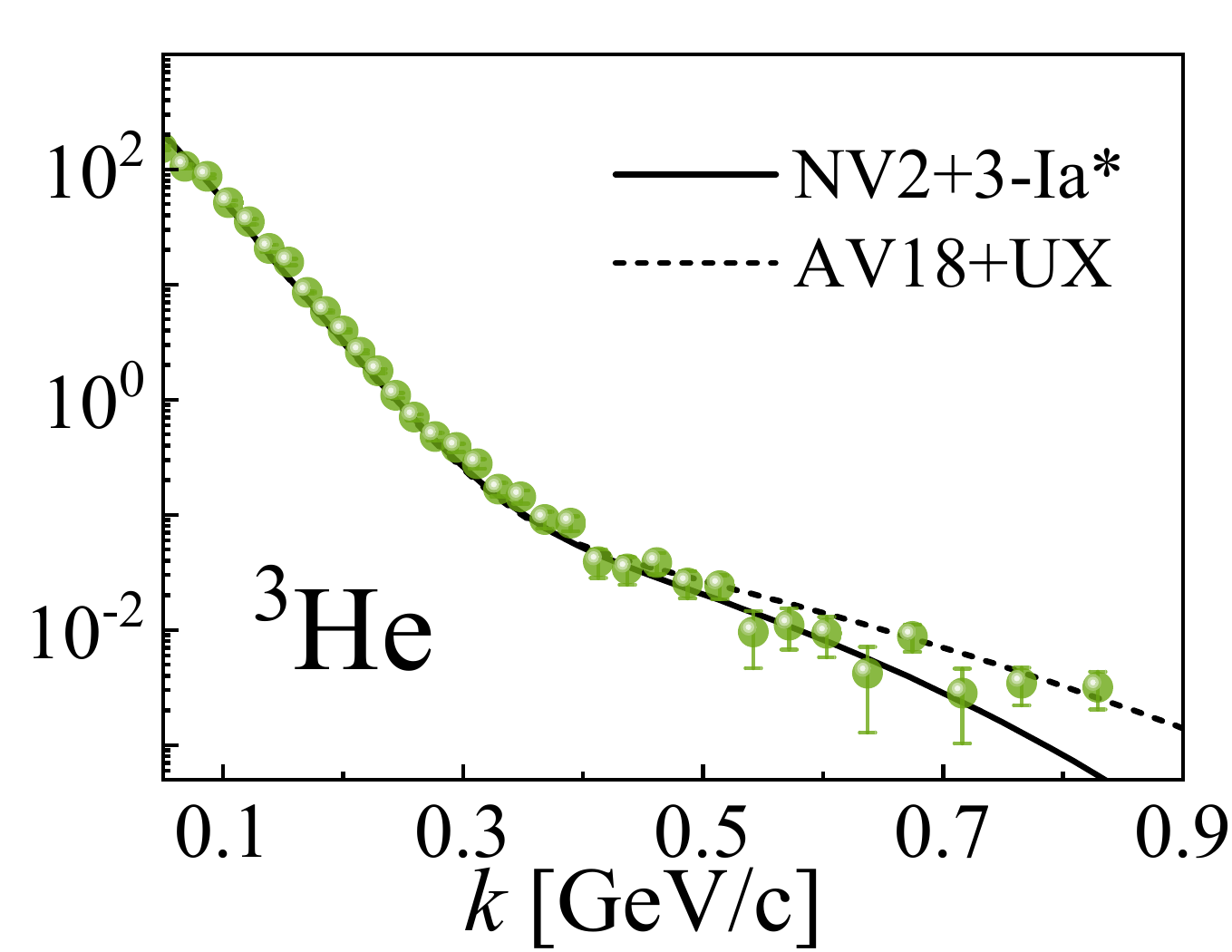}
\end{subfigure}
\hfill
\begin{subfigure}{0.313\textwidth}
  \caption{($5.766~\text{GeV},18^{\circ}$), $Q^2=2.5~(\text{GeV/c})^2$}
  \label{nk_He4_5766_18}
  \includegraphics[width=\linewidth]{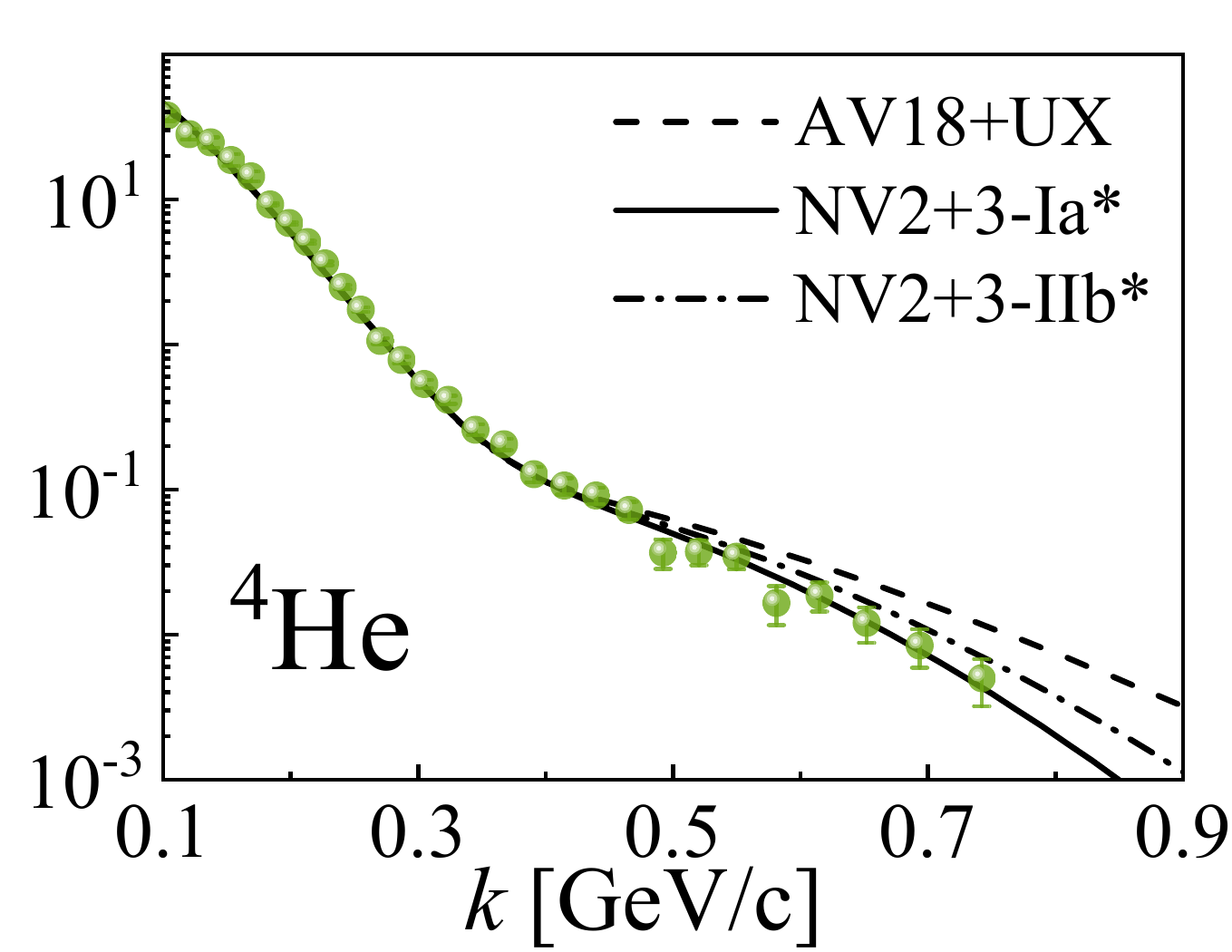}
\end{subfigure}

\vspace{2mm}

\begin{subfigure}{0.333\textwidth}
  \caption{($5.766~\text{GeV},18^{\circ}$), $Q^2=2.5~(\text{GeV/c})^2$}
  \label{nk_Be9_5766_18}
  \includegraphics[width=\linewidth]{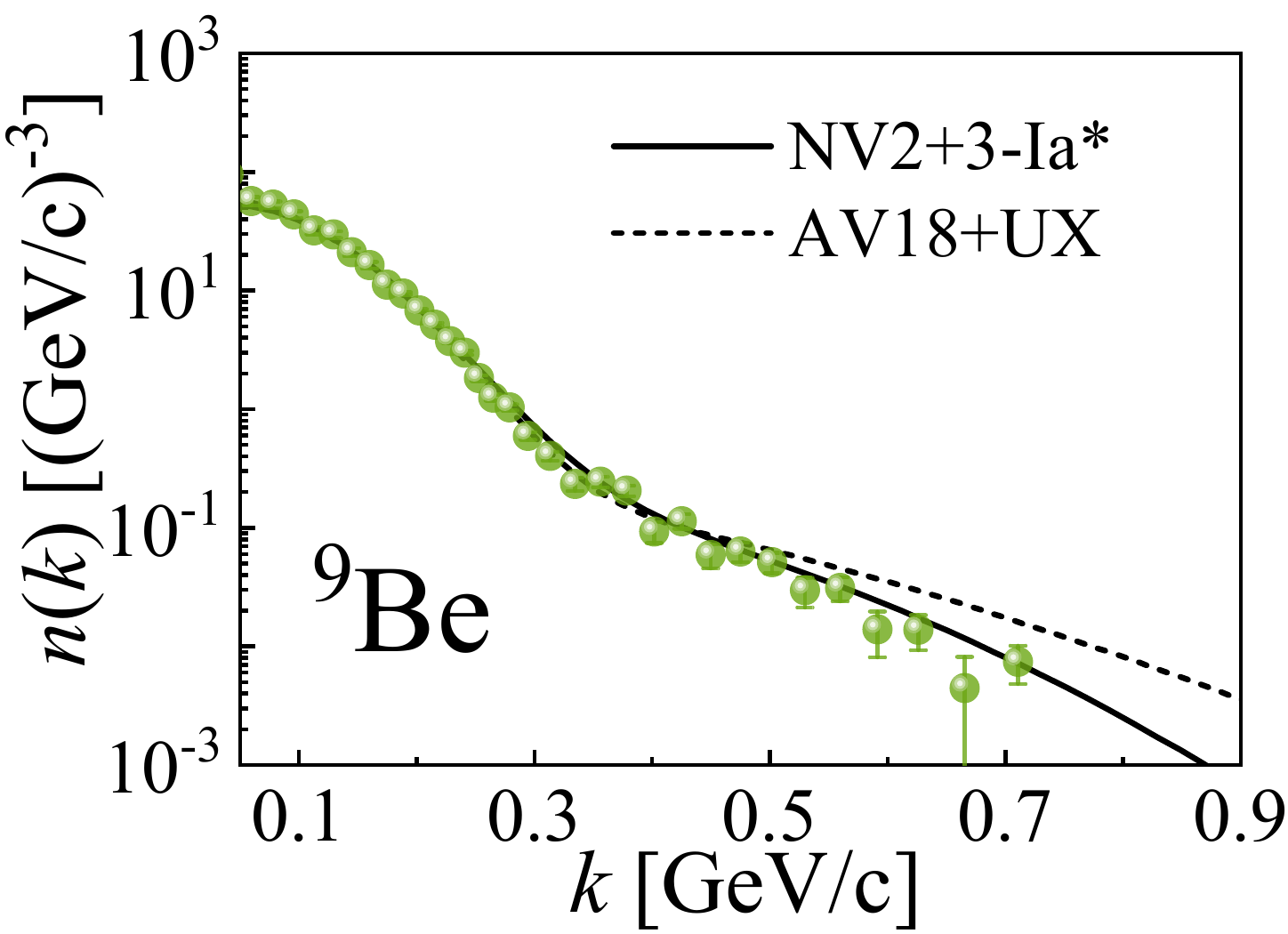}
\end{subfigure}
\hfill
\begin{subfigure}{0.313\textwidth}
  \caption{($5.766~\text{GeV},18^{\circ}$), $Q^2=2.5~(\text{GeV/c})^2$}
  \label{C12_5766_18}
  \includegraphics[width=\linewidth]{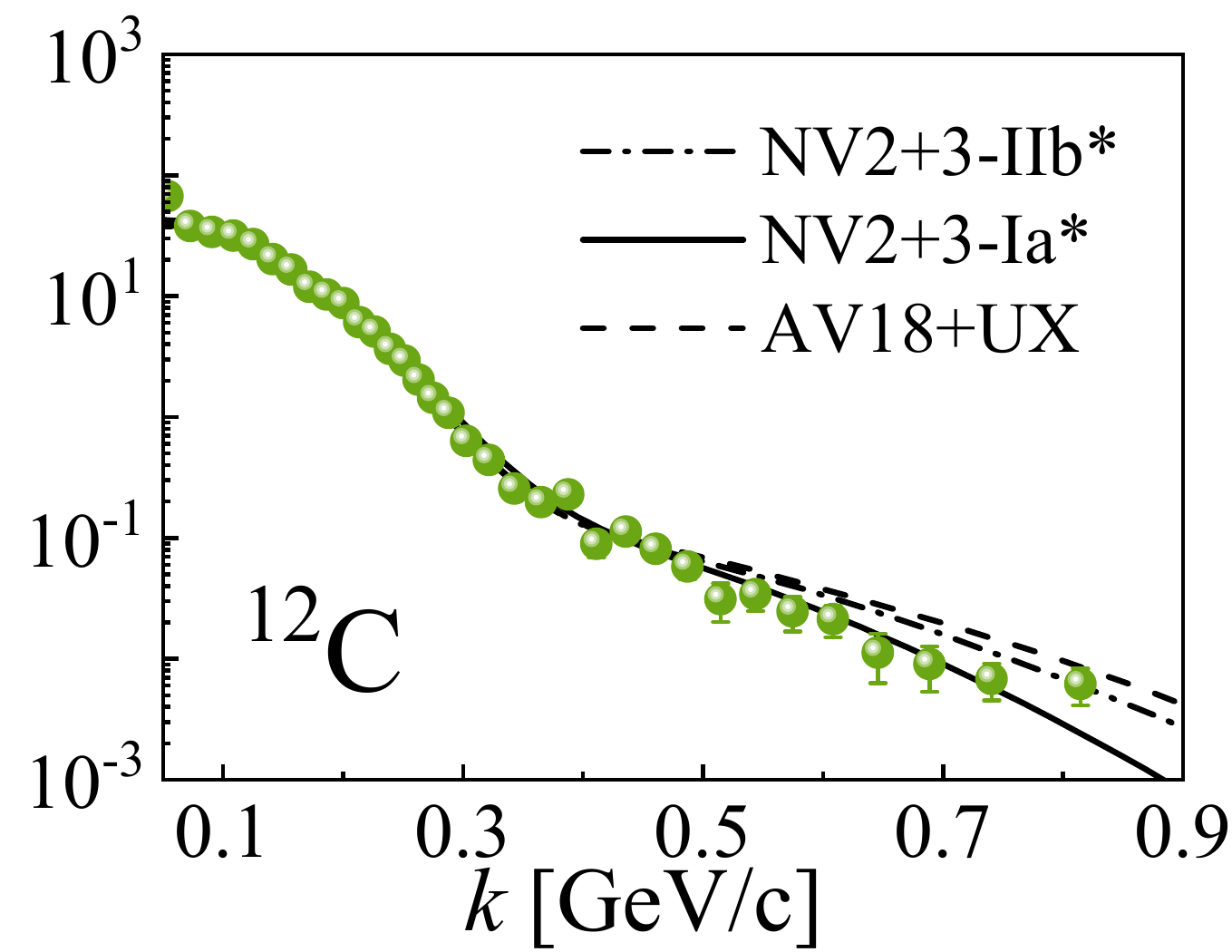}
\end{subfigure}
\hfill
\begin{subfigure}{0.313\textwidth}
  \caption{($3.356~\text{GeV},25^{\circ}$), $Q^2=1.6~(\text{GeV/c})^2$}
  \label{nk_Ca40_3356_25}
  \includegraphics[width=\linewidth]{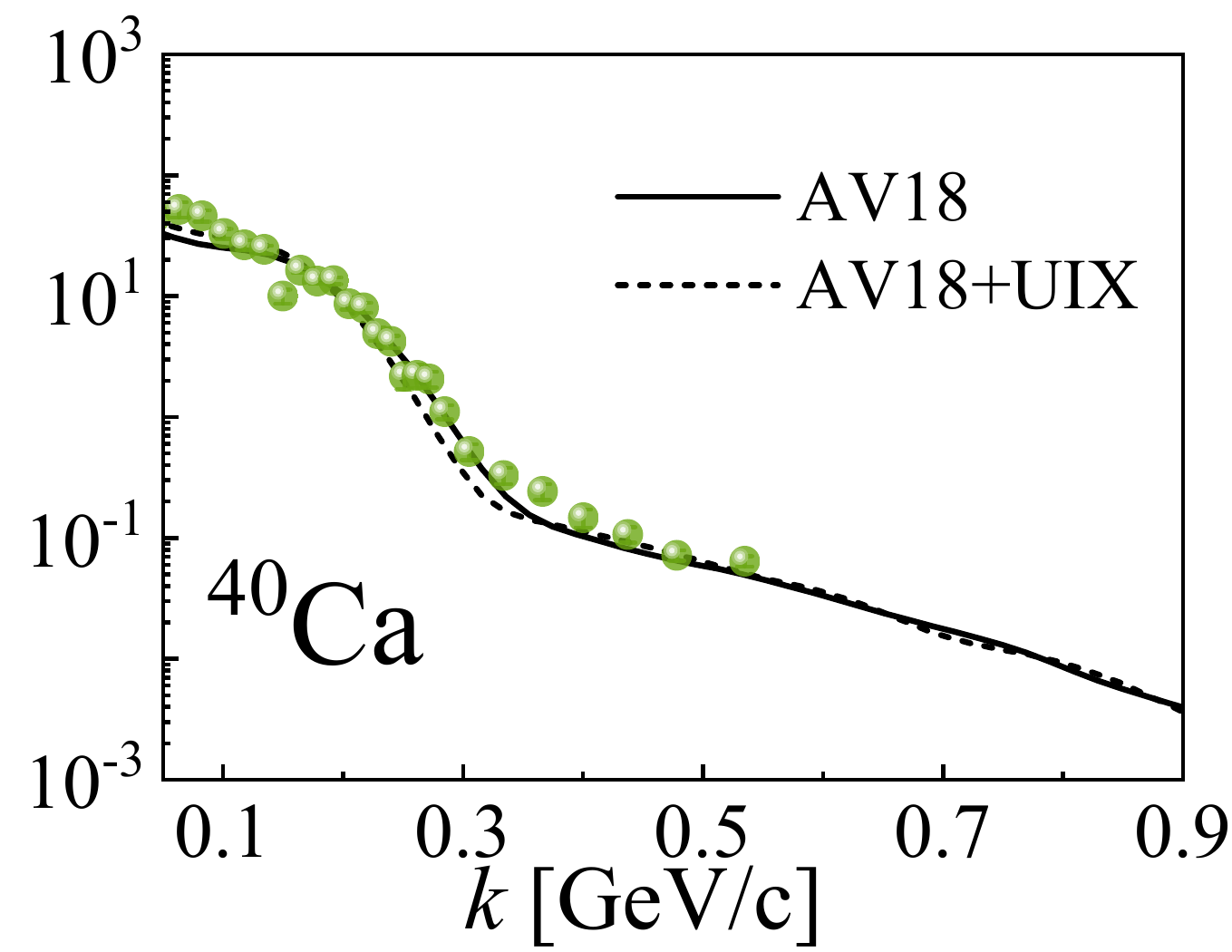}
\end{subfigure}

\vspace{2mm}

\begin{subfigure}{0.333\textwidth}
  \caption{($4.045~\text{GeV},23^{\circ}$), $Q^2=1.9~(\text{GeV/c})^2$}
  \label{nk_Fe56_4045_23}
  \includegraphics[width=\linewidth]{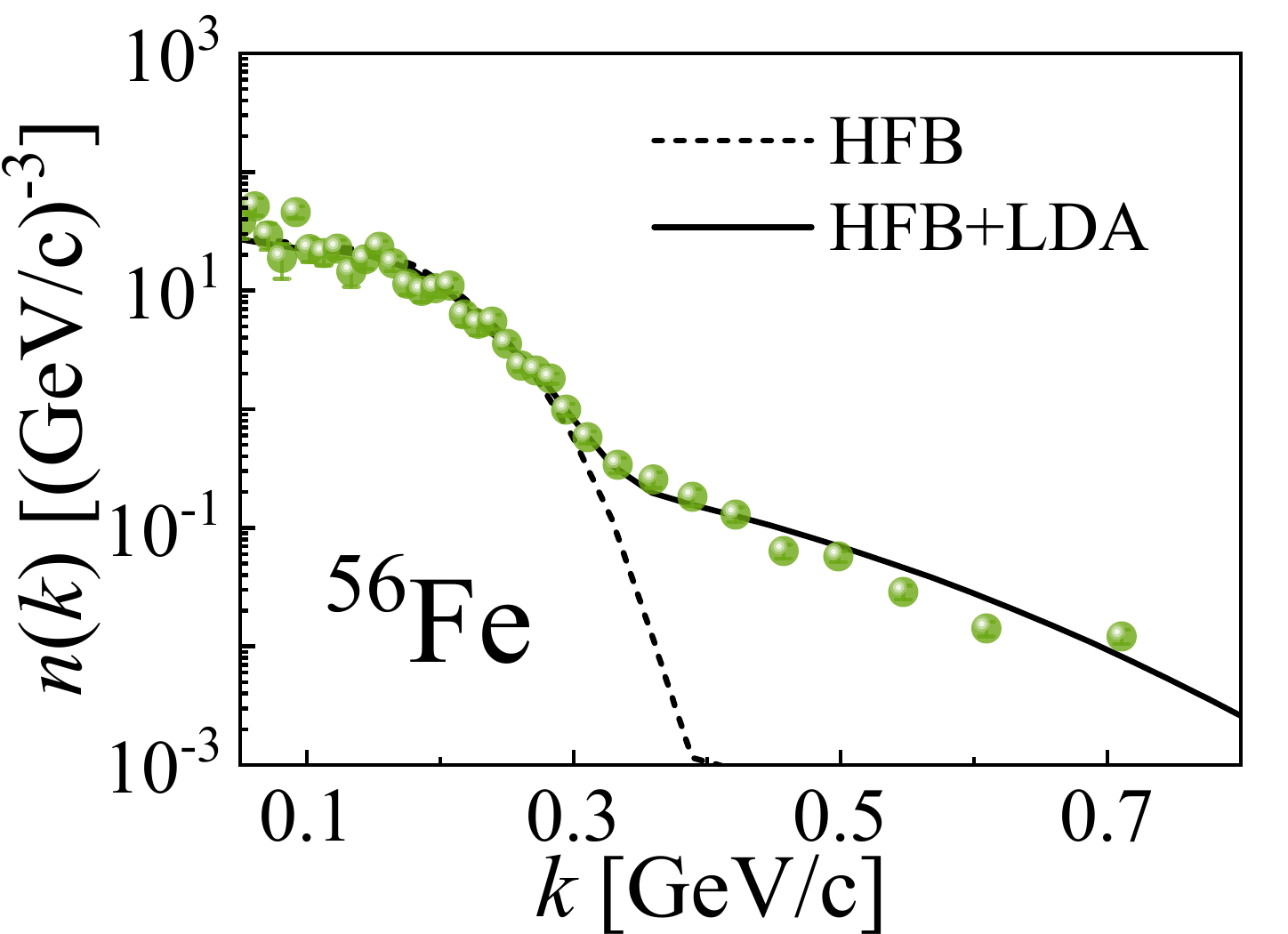}
\end{subfigure}
\hfill
\begin{subfigure}{0.313\textwidth}
  \caption{($4.045~\text{GeV},23^{\circ}$), $Q^2=1.9~(\text{GeV/c})^2$}
  \label{nk_Au197_4045_23}
  \includegraphics[width=\linewidth]{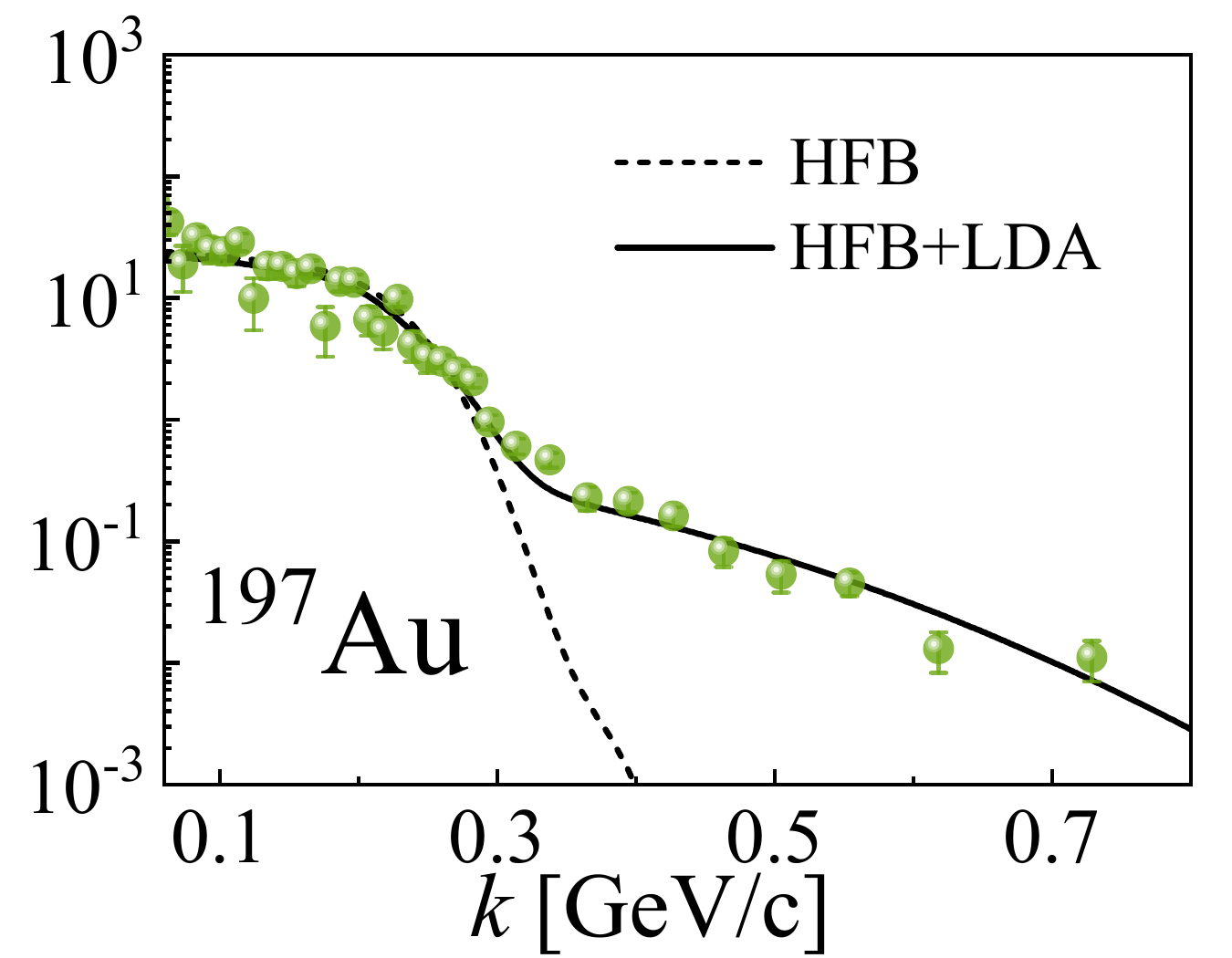}
\end{subfigure}
\hfill
\begin{subfigure}{0.313\textwidth}
  \caption{($3.595~\text{GeV},20^{\circ}$), $Q^2=1.3~(\text{GeV/c})^2$}
  \label{nk_NM_3595_20}
  \includegraphics[width=\linewidth]{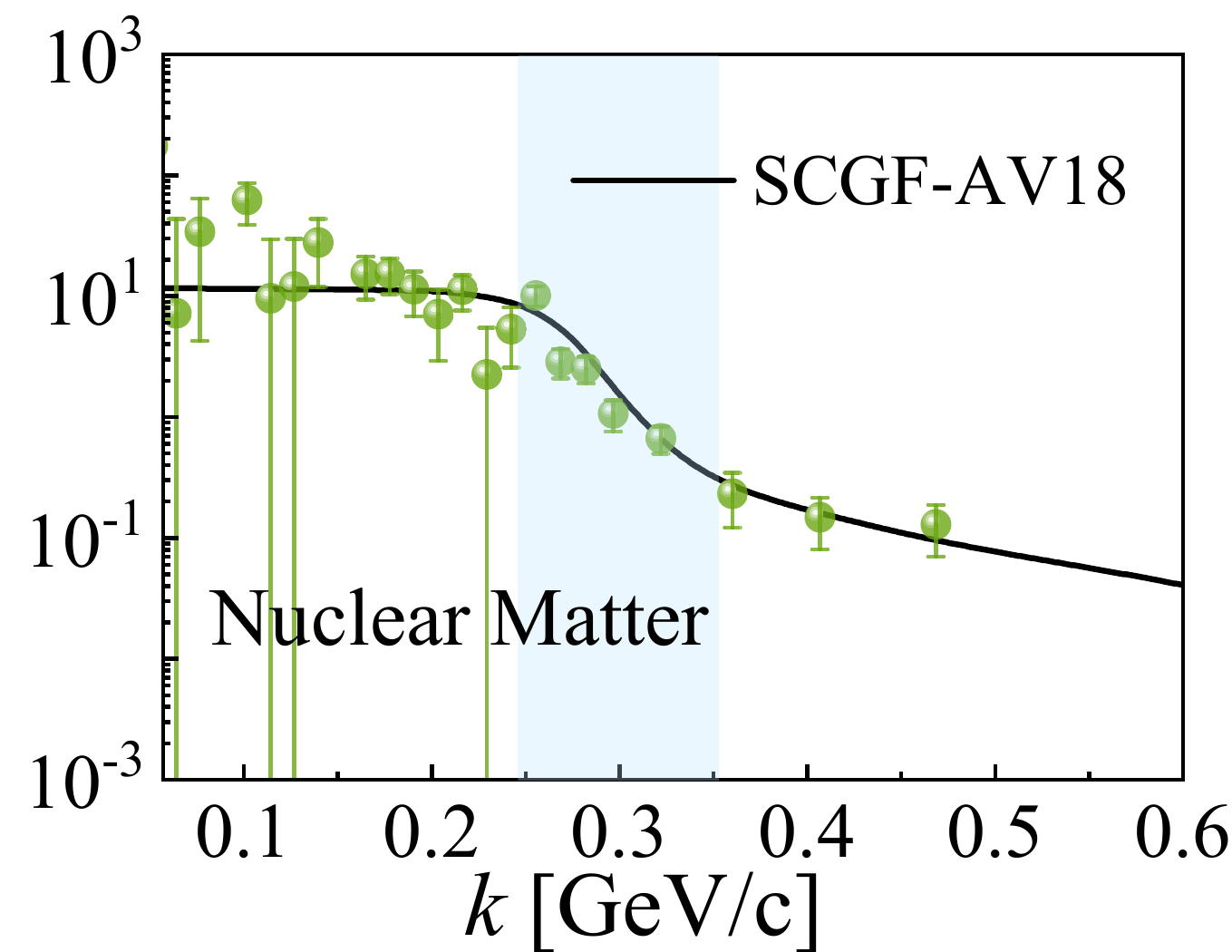}
\end{subfigure}

\caption{
Nucleon momentum distributions of the deuteron, $^{3}$He, $^{4}$He, $^{9}$Be, $^{12}$C, $^{40}$Ca, $^{56}$Fe, $^{197}$Au, and nuclear matter, extracted from inclusive electron scattering data.
The incident electron energy and scattering angle $(E_e,\theta)$, and corresponding $Q^{2}$ values are indicated.
For nuclei with $A\le 40$, the curves show QMC results with different $NN+3N$ interactions~\cite{Wiringa2014}.
For $^{56}$Fe and $^{197}$Au, dashed lines correspond to Hartree-Fock-Bogolyubov (HFB) calculations, and solid lines present correlation results with the local density approximation (LDA)~\cite{Gaidarov2009,Stoitsov2004}.
For nuclear matter, the curve represents self-consistent Green's function (SCGF) results employing the AV18 interaction~\cite{Rios2013}.
The color bands indicate the region $(k_F-0.05~\text{GeV/c},\,k_F+0.05~\text{GeV/c})$, approximately corresponding to the transition from Fermi motion to SRCs.
}
\label{other}
\end{figure*}

Having validated the RFG excitation energy for $^{12}$C and $^{4}$He, Fig.~\ref{other} shows the extracted nucleon momentum distributions $n(k)$ for a broad range of nuclei, including $^{3}$He, $^{4}$He, $^{9}$Be, $^{12}$C, $^{40}$Ca, $^{56}$Fe, $^{197}$Au, and nuclear matter.
The experimental data are taken from Ref.~\cite{Benhar2006}.
For comparison, the $n(k)$ extractions of the deuteron, with no excitation energy, are also provided~\cite{Fomin2012,Liang2024}.  
For nuclei with $A\le 40$, QMC calculations with $NN+3N$ interactions are included for comparison~\cite{Wiringa2014}; for $^{56}$Fe and $^{197}$Au, the theoretical results are given by the Hartree-Fock-Bogolyubov (HFB) model with the high-momentum tails provided by the local-density approximation (LDA)~\cite{Gaidarov2009,Stoitsov2004}.  
For nuclear matter, we present the self-consistent Green’s function (SCGF) results employing the AV18 interaction~\cite{Rios2013}.

Overall, the extracted $n(k)$ of these nuclei exhibits good agreement with theory across both the low- and high-momentum regions. 
Obvious low-momentum plateaus caused by the Fermi motion are observed below the Fermi momentum.
From light to heavy nuclei, the low-momentum plateau becomes increasingly pronounced, and this evolution is well reproduced by the $n(k)$ extractions. 
In particular, for nuclear matter, a well-defined Fermi distribution is clearly observed below $k_F=0.3$ GeV/c.
At high momenta, the extracted $n(k)$ of all these nuclei correctly follows the universal high-momentum tails as observed in the theoretical calculations, reflecting the dominant role of short-range correlations.

Focusing on the transition region around $k_F$, where both the Fermi motion and SRCs contribute significantly to the nucleon momentum distribution, the extracted $n(k)$ results also show good consistency with the theoretical calculations for all these nuclei.
Moreover, the transition from Fermi motion to SRCs, which is expected to sharpen with increasing mass number $A$, is also captured.  
This feature is most evident in nuclear matter, where a distinct transition region around $k \simeq k_F$ (shaded area) is reproduced by the extraction.
These results further confirm that the proposed $E^{*\text{RFG}}_{A-1}$ description can well characterize the excitation energy of the residual $A-1$ system for complex nuclei.

\section{conclusions}
\label{Concl}

In this work, we successfully extract the nucleon momentum distributions $n(k)$ for a variety of complex nuclei from inclusive electron scattering data.
Within the relativistic Fermi gas (RFG) framework, we developed a unified treatment of the excitation energy of the residual system that incorporates both Fermi motion and SRCs.
Introducing this RFG excitation energy resolves the inconsistency between the extracted $n(k)$ and ab initio calculations, especially around the Fermi momentum $k_F$.

For a wide range of complex nuclei from $^{3,4}$He to medium and heavy nuclei, as well as nuclear matter, the extracted $n(k)$ exhibits good agreement with state-of-the-art theoretical calculations.
The resulting momentum distributions clearly display the expected trends: an increasingly pronounced Fermi-motion-induced low-momentum plateau with growing mass number, a universal high-momentum tail shared by all nuclei and nuclear matter, and a well-defined transition from low-momentum plateau to high-momentum tail near $k_F$.
Overall, this work establishes a robust method for describing the excitation energy and extracts reliable experimental nucleon momentum distributions from inclusive scattering data for complex nuclei.
These results offer new insight into the interplay between Fermi motion and short-range correlations in complex nuclei and provide a feasible approach for future studies of energy and momentum dependence in inclusive and exclusive scattering experiments.

	\begin{acknowledgments}
The authors would like to thank Prof. Zhihong Ye in Tsinghua University for valuable suggestions and warmhearted encouragement. This work is supported by the National Natural Science Foundation of China (Grants No.\ 12547176, No.\ 12535009, No.\ 11975167,  No.\ 11947211, No.\ 11905103,  No.\  11881240623, No.\ 11961141003, and No.\ 12375122), and by the National Key R\&D Program of China (Contracts No.\ 2023YFA1606503 and No.\ 2018YFA0404403). 
	\end{acknowledgments}

	\normalem
	\bibliography{references}

@article{Sarriguren2019,
    author = "Sarriguren, P.",
    title = "{Mean-field calculations of charge radii in ground and isomeric states of Cd isotopes}",
    doi = "10.1103/PhysRevC.100.054306",
    journal = "Phys. Rev. C",
    volume = "100",
    number = "5",
    pages = "054306",
    year = "2019"
}

@article{Sarriguren2019tqu,
    author = "Sarriguren, P. and Merino, D. and Moreno, O. and Moya De Guerra, E. and Kadrev, D. N. and Antonov, A. N. and Gaidarov, M. K.",
    title = "{Elastic magnetic electron scattering from deformed nuclei}",
    doi = "10.1103/PhysRevC.99.034325",
    journal = "Phys. Rev. C",
    volume = "99",
    number = "3",
    pages = "034325",
    year = "2019"
}

@article{Reinhard2021,
    author = "Reinhard, Paul-Gerhard and Nazarewicz, Witold",
    title = "{Nuclear charge densities in spherical and deformed nuclei: towards precise calculations of charge radii}",
    doi = "10.1103/PhysRevC.103.054310",
    journal = "Phys. Rev. C",
    volume = "103",
    pages = "054310",
    year = "2021"
}

@article{Wang2021,
   author = {Wang, X. and Niu, Q. and Zhang, J. and Lyu, M. and Liu, J. and Xu, C. and Ren, Z.},
   title = {Nucleon momentum distribution of Fe-56 from the axially deformed relativistic mean-field model with nucleon-nucleon correlations},
   journal = {Sci. China Phys. Mech. Astron.},
   volume = {64},
pages ={292011},
   number = {9},
   ISSN = {1674-7348},
   url = {https://doi.org/10.1007/s11433-021-1729-5},
   year = {2021},
   type = {Journal Article}
}

@article{MoyadeGuerra1991,
    author = "Moya de Guerra, E. and Sarriguren, P. and Caballero, J. A. and Casas, M. and Sprung, D. W. L.",
    title = "{Momentum distributions from deformed Hartree-Fock calculations}",
    doi = "10.1016/0375-9474(91)90786-6",
    journal = "Nucl. Phys. A",
    volume = "529",
    pages = "68--94",
    year = "1991"
}

@article{Stoitsov2004,
    author = "Stoitsov, M. V. and Dobaczewski, J. and Nazarewicz, W. and Ring, P.",
    title = "{Axially deformed solution of the Skyrme-Hartree-Fock-Bogolyubov equations using the transformed harmonic oscillator basis: The Program HFBTHO (v1.66p)}",
    doi = "10.1016/j.cpc.2005.01.001",
    journal = "Comput. Phys. Commun.",
    volume = "167",
    pages = "43--63",
    year = "2005"
}

@article{SCHUNCK2017,
title = {Solution of the Skyrme-Hartree–Fock–Bogolyubovequations in the Cartesian deformed harmonic-oscillator basis. (VIII) hfodd (v2.73y): A new version of the program},
journal = {Comput. Phys. Commun.},
volume = {216},
pages = {145-174},
year = {2017},
issn = {0010-4655},
doi = {https://doi.org/10.1016/j.cpc.2017.03.007},
url = {https://www.sciencedirect.com/science/article/pii/S0010465517300942},
author = {N. Schunck and J. Dobaczewski and W. Satuła and P. Bączyk and J. Dudek and Y. Gao and M. Konieczka and K. Sato and Y. Shi and X.B. Wang and T.R. Werner}
}

@article{Li2025,
    author = "Pengcheng Li and Manzi Nan and Haojie Zhang and Junhuai Xu and Xilong Xiang and Yijie Wang and Yongjia Wang and Gaochan Yong and Tadaaki Isobe and Zhigang Xiao and Qingfeng Li",
    title = "{Unlocking the initial neutron density distribution from the two-pion HBT correlation function in heavy-ion collisions}",
    doi = "10.1016/j.physletb.2025.139963",
    journal = "Phys. Lett. B",
    volume = "870",
    pages = "139963",
    year = "2025"
}

@article{Qin2023,
    author = "Qin, Yuhao and others",
    title = "{Probing high-momentum component in nucleon momentum distribution by neutron-proton bremsstrahlung {\ensuremath{\gamma}}-rays in heavy ion reactions}",
    doi = "10.1016/j.physletb.2024.138514",
    journal = "Phys. Lett. B",
    volume = "850",
    pages = "138514",
    year = "2024"
}

@article{Niu2022,
  title = {Effects of nucleon-nucleon short-range correlations on inclusive electron scattering},
  author = {Niu, Qinglin and Liu, Jian and Guo, Yuanlong and Xu, Chang and Lyu, Mengjiao and Ren, Zhongzhou},
  journal = {Phys. Rev. C},
  volume = {105},
  issue = {5},
  pages = {L051602},
  numpages = {6},
  year = {2022},
  month = {May},
  publisher = {American Physical Society},
  doi = {10.1103/PhysRevC.105.L051602},
  url = {https://link.aps.org/doi/10.1103/PhysRevC.105.L051602}
}

@article{Duer2019,
  title = {Direct Observation of Proton-Neutron Short-Range Correlation Dominance in Heavy Nuclei},
  author = {Duer, M. and others},
  collaboration = {CLAS Collaboration},
  journal = {Phys. Rev. Lett.},
  volume = {122},
  issue = {17},
  pages = {172502},
  numpages = {8},
  year = {2019},
  month = {May},
  publisher = {American Physical Society},
  doi = {10.1103/PhysRevLett.122.172502},
  url = {https://link.aps.org/doi/10.1103/PhysRevLett.122.172502}
}

@article{Hen2014,
    author = "Hen, O. and others",
    title = "{Momentum sharing in imbalanced Fermi systems}",
    doi = "10.1126/science.1256785",
    journal = "Science",
    volume = "346",
    pages = "614--617",
    year = "2014"
}

@article{Cruz-Torres2019,
    author = "Cruz-Torres, R. and Lonardoni, D. and Weiss, R. and Barnea, N. and Higinbotham, D. W. and Piasetzky, E. and Schmidt, A. and Weinstein, L. B. and Wiringa, R. B. and Hen, O.",
    title = "{Many-body factorization and position{\textendash}momentum equivalence of nuclear short-range correlations}",
    reportNumber = "LA-UR-19-25832",
    doi = "10.1038/s41567-020-01053-7",
    journal = "Nature Phys.",
    volume = "17",
    number = "3",
    pages = "306--310",
    year = "2021"
}

@article{Weiss2016,
    author = "Weiss, R. and Cruz-Torres, R. and Barnea, N. and Piasetzky, E. and Hen, O.",
    title = "{The nuclear contacts and short range correlations in nuclei}",
    doi = "10.1016/j.physletb.2018.01.061",
    journal = "Phys. Lett. B",
    volume = "780",
    pages = "211--215",
    year = "2018"
}

@article{Tropiano2021,
    author = "Tropiano, A. J. and Bogner, S. K. and Furnstahl, R. J.",
    title = "{Short-range correlation physics at low renormalization group resolution}",
    doi = "10.1103/PhysRevC.104.034311",
    journal = "Phys. Rev. C",
    volume = "104",
    number = "3",
    pages = "034311",
    year = "2021"
}

@article{Tropiano2024,
    author = "Tropiano, A. J. and Bogner, S. K. and Furnstahl, R. J. and Hisham, M. A. and Lovato, A. and Wiringa, R. B.",
    title = "{High-resolution momentum distributions from low-resolution wave functions}",
    doi = "10.1016/j.physletb.2024.138591",
    journal = "Phys. Lett. B",
    volume = "852",
    pages = "138591",
    year = "2024"
}

@article{Wiringa2014,
  title = {Nucleon and nucleon-pair momentum distributions in $A\ensuremath{\le}12$ nuclei},
  author = {Wiringa, R. B. and Schiavilla, R. and Pieper, Steven C. and Carlson, J.},
  journal = {Phys. Rev. C},
  volume = {89},
  issue = {2},
  pages = {024305},
  numpages = {9},
  year = {2014},
  month = {Feb},
  publisher = {American Physical Society},
  doi = {10.1103/PhysRevC.89.024305},
  url = {https://link.aps.org/doi/10.1103/PhysRevC.89.024305}
}

@article{Piarulli2022,
    author = "Piarulli, M. and Pastore, S. and Wiringa, R. B. and Brusilow, S. and Lim, R.",
    title = "{Densities and momentum distributions in A{\ensuremath{\leq}}12 nuclei from chiral effective field theory interactions}",
    doi = "10.1103/PhysRevC.107.014314",
    journal = "Phys. Rev. C",
    volume = "107",
    number = "1",
    pages = "014314",
    year = "2023"
}

@article{Marcucci2019,
  title = {Momentum distributions and short-range correlations in the deuteron and $^{3}\mathrm{He}$ with modern chiral potentials},
  author = {Marcucci, Laura Elisa and Sammarruca, Francesca and Viviani, Michele and Machleidt, Ruprecht},
  journal = {Phys. Rev. C},
  volume = {99},
  issue = {3},
  pages = {034003},
  numpages = {18},
  year = {2019},
  month = {Mar},
  publisher = {American Physical Society},
  doi = {10.1103/PhysRevC.99.034003},
  url = {https://link.aps.org/doi/10.1103/PhysRevC.99.034003}
}

@article{Neff2015,
  title = {Short-range correlations in nuclei with similarity renormalization group transformations},
  author = {Neff, T. and Feldmeier, H. and Horiuchi, W.},
  journal = {Phys. Rev. C},
  volume = {92},
  issue = {2},
  pages = {024003},
  numpages = {12},
  year = {2015},
  month = {Aug},
  publisher = {American Physical Society},
  doi = {10.1103/PhysRevC.92.024003},
  url = {https://link.aps.org/doi/10.1103/PhysRevC.92.024003}
}

@article{Ciofi1991,
    author = "Ciofi degli Atti, Claudio and Pace, E. and Salme, G.",
    title = "{Y scaling analysis of quasielastic electron scattering and nucleon momentum distributions in few body systems, complex nuclei and nuclear matter}",
    reportNumber = "INFN-ISS-90-8",
    doi = "10.1103/PhysRevC.43.1155",
    journal = "Phys. Rev. C",
    volume = "43",
    pages = "1155--1176",
    year = "1991"
}

@article{Wang2023,
    author = "Wang, Lei and Niu, Qinglin and Zhang, Jinjuan and Liu, Jian and Ren, Zhongzhou",
    title = "{New extended method for \ensuremath{\psi}' scaling function of inclusive electron scattering}",
    doi = "10.1007/s11433-023-2135-x",
    journal = "Sci. China Phys. Mech. Astron.",
    volume = "66",
    number = "10",
    pages = "102011",
    year = "2023"
}

@article{Liang2022,
  title = {Nucleon momentum distributions from inclusive electron scattering with superscaling analysis},
  author = {Liang, Tongqi and Ren, Zhongzhou and Bai, Dong and Liu, Jian},
  journal = {Phys. Rev. C},
  volume = {106},
  issue = {5},
  pages = {054324},
  numpages = {7},
  year = {2022},
  month = {Nov},
  publisher = {American Physical Society},
  doi = {10.1103/PhysRevC.106.054324},
  url = {https://link.aps.org/doi/10.1103/PhysRevC.106.054324}
}

@article{Fomin2012,
   author = {Fomin, N. and others},
   title = {New Measurements of High-Momentum Nucleons and Short-Range Structures in Nuclei},
   journal = {Phys. Rev. Lett.},
   volume = {108},
   number = {9},
   pages = {092502},
   DOI = {10.1103/PhysRevLett.108.092502},
   url = {https://link.aps.org/doi/10.1103/PhysRevLett.108.092502},
   year = {2012},
   type = {Journal Article}
}

@article{Arrington2022,
author = {Arrington, John and Fomin, Nadia and Schmidt, Axel},
title = {Progress in Understanding Short-Range Structure in Nuclei: An Experimental Perspective},
journal = {Annu. Rev. Nucl. Part. Sci.},
volume = {72},
number = {1},
pages = {307-337},
year = {2022},
doi = {10.1146/annurev-nucl-102020-022253}
}

@article{Ciofi1999,
    author = "Ciofi degli Atti, Claudio and West, Geoffrey B.",
    title = "{A New approach to y scaling and the universal features of scaling functions and nucleon momentum distributions}",
    doi = "10.1016/S0370-2693(99)00599-7",
    journal = "Phys. Lett. B",
    volume = "458",
    pages = "447--453",
    year = "1999"
}

@article{Ciofi2015,
    author = "Ciofi degli Atti, Claudio",
    title = "{In-medium short-range dynamics of nucleons: Recent theoretical and experimental advances}",
    doi = "10.1016/j.physrep.2015.06.002",
    journal = "Phys. Rept.",
    volume = "590",
    pages = "1--85",
    year = "2015"
}

@article{Caballero2010,
   author = {Caballero, J. A. and Barbaro, M. B. and Antonov, A. N. and Ivanov, M. V. and Donnelly, T. W.},
   title = {Scaling function and nucleon momentum distribution},
   journal = {Phys. Rev. C},
   volume = {81},
   number = {5},
   pages = {055502},
   DOI = {10.1103/PhysRevC.81.055502},
   url = {https://link.aps.org/doi/10.1103/PhysRevC.81.055502},
   year = {2010},
   type = {Journal Article}
}

@article{Day1990,
   author = {Day, D. B. and McCarthy, J. S. and Donnelly, T. W. and Sick, I},
   title = {Scaling in inclusive electron-nucleus scattering},
   journal = {Annu. Rev. Nucl. Part. Sci.},
   volume = {40},
   number = {1},
   pages = {357-410},
   ISSN = {0163-8998},
   year = {1990},
   URL = { 
        https://doi.org/10.1146/annurev.ns.40.120190.002041
},
   type = {Journal Article}
}

@article{Ivanov2020,
    author = "Ivanov, M. V. and Antonov, A. N. and Caballero, J. A.",
    title = "{Nucleon momentum distribution extracted from the experimental scaling function}",
    doi = "10.1016/j.nuclphysa.2020.122029",
    journal = "Nucl. Phys. A",
    volume = "1003",
    pages = "122029",
    year = "2020"
}

@article{Meng2023,
    author = "Meng, Qi and Lu, Ziyang and Xu, Chang",
    title = "{Short-range correlations and momentum distributions in mirror nuclei H3 and He3}",
    doi = "10.1103/PhysRevC.108.014001",
    journal = "Phys. Rev. C",
    volume = "108",
    number = "1",
    pages = "014001",
    year = "2023"
}

@article{Li2022,
    author = "Li, S. and Cruz-Torres, R and Santiesteban, N and Ye, Z and others",
    title = "{Revealing the short-range structure of the mirror nuclei $^{3}$H and $^{3}$He}",
    doi = "10.1038/s41586-022-05007-2",
    journal = "Nature",
    volume = "609",
    number = "7925",
    pages = "41--45",
    year = "2022"
}

@article{Li2024,
    author = "Li, S. and others",
    title = "{Inclusive studies of two- and three-nucleon short-range correlations in 3H and 3He}",
    reportNumber = "JLAB-PHY-25-4410",
    doi = "10.1016/j.physletb.2025.139734",
    journal = "Phys. Lett. B",
    volume = "868",
    pages = "139734",
    year = "2025"
}

@article{Schmidt2024,
    author = "Schmidt, A. and Denniston, A. W. and Seroka, E. M. and Barnea, N. and Higinbotham, D. W. and Korover, I. and Miller, G. A. and Piasetzky, E. and Strikman, M. and Weinstein, L. B. and Weiss, R. and Hen, O.",
    title = "{A=3(e,e')xB{\ensuremath{\geq}}1 cross-section~ratios and the isospin structure of short-range correlations}",
    doi = "10.1103/PhysRevC.109.054001",
    journal = "Phys. Rev. C",
    volume = "109",
    number = "5",
    pages = "054001",
    year = "2024"
}

@article{Zhang2025,
title = {Measuring short-range correlations and quasi-elastic cross sections in A(e,e’) at x > 1 and modest Q2},
journal = {Phys. Lett. B},
pages = {140087},
year = {2025},
issn = {0370-2693},
doi = {https://doi.org/10.1016/j.physletb.2025.140087},
url = {https://www.sciencedirect.com/science/article/pii/S0370269325008457},
author = {Y.P. Zhang and Z.H. Ye and D. Nguyen and P. Aguilera and Z. Ahmed and others}
}

@article{Ciofi2009,
    author = "Ciofi degli Atti, Claudio and Mezzetti, Chiara Benedetta",
    title = "{Obtaining information on Short Range Correlations from inclusive electron scattering}",
    doi = "10.1103/PhysRevC.79.051302",
    journal = "Phys. Rev. C",
    volume = "79",
    pages = "051302",
    year = "2009"
}

@article{Benhar2013,
    author = "Benhar, Omar",
    title = "{Final state interactions in the nuclear response at large momentum transfer}",
    doi = "10.1103/PhysRevC.87.024606",
    journal = "Phys. Rev. C",
    volume = "87",
    number = "2",
    pages = "024606",
    year = "2013"
}

@article{Ciofi1994,
    author = "Ciofi degli Atti, Claudio and Simula, S.",
    title = "{Nucleon-nucleon correlations and final state interactions in inclusive quasielastic electron scattering off nuclei at x {\ensuremath{>}} 1}",
    reportNumber = "INFN-ISS-94-1",
    doi = "10.1016/0370-2693(94)90010-8",
    journal = "Phys. Lett. B",
    volume = "325",
    pages = "276--282",
    year = "1994"
}

@article{Rocco2015,
    author = "Rocco, Noemi and Lovato, Alessandro and Benhar, Omar",
    title = "{Unified description of electron-nucleus scattering within the spectral function formalism}",
    doi = "10.1103/PhysRevLett.116.192501",
    journal = "Phys. Rev. Lett.",
    volume = "116",
    number = "19",
    pages = "192501",
    year = "2016"
}

@article{Weiss2018,
    author = "Weiss, Ronen and Korover, Igor and Piasetzky, Eliezer and Hen, Or and Barnea, Nir",
    title = "{Energy and momentum dependence of nuclear short-range correlations - Spectral function, exclusive scattering experiments and the contact formalism}",
    doi = "10.1016/j.physletb.2019.02.019",
    journal = "Phys. Lett. B",
    volume = "791",
    pages = "242--248",
    year = "2019"
}

@article{Ankowski2024,
    author = "Ankowski, Artur M. and Benhar, Omar and Sakuda, Makoto",
    title = "{Determination of the proton spectral function of C12 from (e,e'p) data}",
    doi = "10.1103/PhysRevC.110.054612",
    journal = "Phys. Rev. C",
    volume = "110",
    number = "5",
    pages = "054612",
    year = "2024"
}

@article{Rohe2004,
    author = "Rohe, D. and others",
    title = "{Correlated strength in nuclear spectral function}",
    reportNumber = "JLAB-PHY-04-42",
    doi = "10.1103/PhysRevLett.93.182501",
    journal = "Phys. Rev. Lett.",
    volume = "93",
    pages = "182501",
    year = "2004"
}

@article{Benmokhtar2004,
    author = "Benmokhtar, F. and others",
    collaboration = "Jefferson Lab Hall A",
    title = "{Measurement of the He-3(e,e-prime p)pn reaction at high missing energies and momenta}",
    reportNumber = "JLAB-PHY-04-29, DAPNIA-04-212",
    doi = "10.1103/PhysRevLett.94.082305",
    journal = "Phys. Rev. Lett.",
    volume = "94",
    pages = "082305",
    year = "2005"
}

@article{Schmidt2020,
    author = "Schmidt, A. and others",
    collaboration = "CLAS",
    title = "{Probing the core of the strong nuclear interaction}",
    doi = "10.1038/s41586-020-2021-6",
    journal = "Nature",
    volume = "578",
    number = "7796",
    pages = "540--544",
    year = "2020"
}

@article{Gu2020,
    author = "Gu, L. and others",
    collaboration = "Jefferson Lab Hall A",
    title = "{Measurement of the Ar(e,e$^\prime$ p) and Ti(e,e$^\prime$ p) cross sections in Jefferson Lab Hall A}",
    reportNumber = "JLAB-PHY-21-3197",
    doi = "10.1103/PhysRevC.103.034604",
    journal = "Phys. Rev. C",
    volume = "103",
    number = "3",
    pages = "034604",
    year = "2021"
}

@article{Jiang2023,
    author = "Jiang, L. and others",
    collaboration = "Jefferson Lab Hall A",
    title = "{Determination of the titanium spectral function from (e,{\,}e'p) data}",
    doi = "10.1103/PhysRevD.107.012005",
    journal = "Phys. Rev. D",
    volume = "107",
    number = "1",
    pages = "012005",
    year = "2023"
}

@article{Liang2024,
    author = "Liang, Tongqi and Bai, Dong and Ren, Zhongzhou",
    title = "{Modern numerical differentiation technique for extracting nucleon momentum distributions}",
    doi = "10.1103/PhysRevC.109.054313",
    journal = "Phys. Rev. C",
    volume = "109",
    number = "5",
    pages = "054313",
    year = "2024"
}

@article{Arrington2007,
    author = "Arrington, J. and Melnitchouk, W. and Tjon, J. A.",
    title = "{Global analysis of proton elastic form factor data with two-photon exchange corrections}",
    reportNumber = "JLAB-THY-07-678",
    doi = "10.1103/PhysRevC.76.035205",
    journal = "Phys. Rev. C",
    volume = "76",
    pages = "035205",
    year = "2007"
}

@article{Kelly2004,
    author = "Kelly, J. J.",
    title = "{Simple parametrization of nucleon form factors}",
    doi = "10.1103/PhysRevC.70.068202",
    journal = "Phys. Rev. C",
    volume = "70",
    pages = "068202",
    year = "2004"
}

@article{Forest1983,
    author = "De Forest, T.",
    title = "{Off-Shell electron Nucleon Cross-Sections. The Impulse Approximation}",
    doi = "10.1016/0375-9474(83)90124-0",
    journal = "Nucl. Phys. A",
    volume = "392",
    pages = "232--248",
    year = "1983"
}

@article{Shneor2007,
    author = "Shneor, R. and others",
    collaboration = "Jefferson Lab Hall A",
    title = "{Investigation of proton-proton short-range correlations via the C-12(e, e-prime pp) reaction}",
    reportNumber = "JLAB-PHY-07-624",
    doi = "10.1103/PhysRevLett.99.072501",
    journal = "Phys. Rev. Lett.",
    volume = "99",
    pages = "072501",
    year = "2007"
}

@article{Arrington1999,
   author = {Arrington, J and others},
   title = {Inclusive electron-nucleus scattering at large momentum transfer},
  journal = {Phys. Rev. Lett.},
  volume = {82},
  issue = {10},
  pages = {2056--2059},
  numpages = {0},
  year = {1999},
  month = {Mar},
  publisher = {American Physical Society},
  doi = {10.1103/PhysRevLett.82.2056},
  url = {https://link.aps.org/doi/10.1103/PhysRevLett.82.2056}
}

@article{Ye2013,
      author={Zhihong Ye},
      year={2014},
      journal={arXiv:1408.5861 [nucl-ex]},
      url={https://arxiv.org/abs/1408.5861}, 
}

@article{Gaidarov2009,
    author = "Gaidarov, M. K. and Krumova, G. Z. and Sarriguren, P. and Antonov, A. N. and Ivanov, M. V. and Moya de Guerra, E.",
    title = "{Momentum distributions in medium and heavy exotic nuclei}",
    doi = "10.1103/PhysRevC.80.054305",
    journal = "Phys. Rev. C",
    volume = "80",
    pages = "054305",
    year = "2009"
}

@article{Rios2013,
    author = "Rios, A. and Polls, A. and Dickhoff, W. H.",
    title = "{Density and isospin asymmetry dependence of high-momentum components}",
    doi = "10.1103/PhysRevC.89.044303",
    journal = "Phys. Rev. C",
    volume = "89",
    number = "4",
    pages = "044303",
    year = "2014"
}

@article{Hen2012,
  title = {New data strengthen the connection between short range correlations and the EMC effect},
  author = {Hen, O. and Piasetzky, E. and Weinstein, L. B.},
  journal = {Phys. Rev. C},
  volume = {85},
  issue = {4},
  pages = {047301},
  numpages = {4},
  year = {2012},
  month = {Apr},
  publisher = {American Physical Society},
  doi = {10.1103/PhysRevC.85.047301},
  url = {https://link.aps.org/doi/10.1103/PhysRevC.85.047301}
}

@article{Hen2016,
    author = "Hen, O. and Miller, G. A. and Piasetzky, E. and Weinstein, L. B.",
    title = "{Nucleon-Nucleon Correlations, Short-lived Excitations, and the Quarks Within}",
    doi = "10.1103/RevModPhys.89.045002",
    journal = "Rev. Mod. Phys.",
    volume = "89",
    number = "4",
    pages = "045002",
    year = "2017"
}

@article{Schmookler2019,
  title={Modified structure of protons and neutrons in correlated pairs},
  journal={Nature},
author ={Schmookler, B. and others},
collaboration = {CLAS Collaboration},
  volume={566},
  number={7744},
  pages={354--358},
  year={2019},
 DOI = {https://doi.org/10.1038/s41586-019-0925-9},
  publisher={Nature Publishing Group UK London}
}

@article{Hong2024,
    author = "Hong, Bin and Liang, Tongqi and Ren, Zhongzhou",
    title = "{Nucleon-nucleon short-range correlations: A hidden driver in binary neutron star inspiral gravitational waves}",
    doi = "10.1016/j.physletb.2024.139225",
    journal = "Phys. Lett. B",
    volume = "860",
    pages = "139225",
    year = "2025"
}

@article{Liang2025,
    author = "Liang, Tongqi and Bai, Dong and Ren, Zhongzhou",
    title = "{Universal laws for nuclear contacts}",
    doi = "10.1016/j.physletb.2025.139351",
    journal = "Phys. Lett. B",
    volume = "863",
    pages = "139351",
    year = "2025"
}

@article{Liang20242,
    author = "Liang, Tongqi and Bai, Dong and Ren, Zhongzhou",
    title = "{Nuclear contacts of unstable nuclei}",
    doi = "10.1016/j.physletb.2024.138965",
    journal = "Phys. Lett. B",
    volume = "857",
    pages = "138965",
    year = "2024"
}

@article{Li20252,
    author = "Li, Pei and Sun, Kai-Jia and Zhou, Bo and Ma, Guo-Liang",
    title = "{Unmasking short-range correlations via initial-state fluctuations in relativistic heavy-ion collisions}",
    journal = "arXiv:2511.23293 [nucl-th]",
    year = "2025",
    url="
https://doi.org/10.48550/arXiv.2511.23293"
}

@article{Cai2025,
    author = "Cai, Bao-Jun and Li, Bao-An and Ma, Yu-Gang",
    title = "{Nucleon Short-Range Correlations and High-Momentum Dynamics: Implications on the Equation of State of Dense Matter}",
    journal = "arXiv:2512.04206 [nucl-th]",
    year = "2025",
    url="
https://doi.org/10.48550/arXiv.2512.04206"
}

@article{Lu2022,
    author = "Lu, Hao and Ren, Zhongzhou and Bai, Dong",
    title = "{Neutron-neutron short-range correlations and their impacts on neutron stars}",
    doi = "10.1016/j.nuclphysa.2022.122408",
    journal = "Nucl. Phys. A",
    volume = "1021",
    pages = "122408",
    year = "2022"
}

@article{Chen2017,
    author = "Chen, Jiunn-Wei and Detmold, William and Lynn, Joel E. and Schwenk, Achim",
    title = "{Short Range Correlations and the EMC Effect in Effective Field Theory}",
    reportNumber = "MIT-CTP-4798",
    doi = "10.1103/PhysRevLett.119.262502",
    journal = "Phys. Rev. Lett.",
    volume = "119",
    number = "26",
    pages = "262502",
    year = "2017"
}

@article{Jokiniemi2021,
    author = "Jokiniemi, Lotta and Soriano, Pablo and Men{\'e}ndez, Javier",
    title = "{Impact of the leading-order short-range nuclear matrix element on the neutrinoless double-beta decay of medium-mass and heavy nuclei}",
    doi = "10.1016/j.physletb.2021.136720",
    journal = "Phys. Lett. B",
    volume = "823",
    pages = "136720",
    year = "2021"
}

@article{Weiss2022,
    author = "Weiss, Ronen and Soriano, Pablo and Lovato, Alessandro and Menendez, Jaview and Wiringa, R. B.",
    title = "{Neutrinoless double-{\ensuremath{\beta}} decay: Combining quantum Monte Carlo and the nuclear shell model with the generalized contact formalism}",
    reportNumber = "LA-UR-21-32144",
    doi = "10.1103/PhysRevC.106.065501",
    journal = "Phys. Rev. C",
    volume = "106",
    number = "6",
    pages = "065501",
    year = "2022"
}

@article{Benhar2006,
    author = "Benhar, Omar and Day, Donal and Sick, Ingo",
    title = "{Inclusive quasi-elastic electron-nucleus scattering}",
    doi = "10.1103/RevModPhys.80.189",
    journal = "Rev. Mod. Phys.",
    volume = "80",
    pages = "189--224",
    year = "2008"
}
\end{document}